\documentclass[aps,prl,twocolumn,groupedaddress,showpacs,dvipdfmx]{revtex4-1}
\usepackage{graphicx}
\usepackage{dcolumn}
\usepackage{bm}
\usepackage{amsmath}
\usepackage{amssymb}
\usepackage{txfonts}
\usepackage{url}
\usepackage[usenames,dvipsnames]{color}
\usepackage{comment}
\definecolor{Gray}{gray}{0.0}
\definecolor{lightGray}{gray}{0.35}

\begin{document}
\title{Phonon dispersions and Fermi surfaces nesting explaining the variety of
  charge ordering in titanium-oxypnictides superconductors}
\author{Kousuke Nakano$^{1}$}
\author{Kenta Hongo$^{1}$}
\author{Ryo Maezono$^{1}$}
\affiliation{$^{1}$
  School of Information Science, JAIST, Asahidai 1-1, Nomi, Ishikawa
  923-1292, Japan. Correspondence and requests for materials should 
  be addressed to K.N. (email:kousuke\_1123@icloud.com) or 
  R.M. (email:rmaezono@mac.com)
}

\date{\today}
\begin{abstract}
\noindent
There has been a puzzle between experiments
and theoretical predictions 
on the charge ordering of layered titanium-oxypnictides superconductors.
Unconventional mechanisms to explain this discrepancy have been
argued so far, even affecting the understanding of superconductivity
on the compound.
We provide a new theoretical prediction, by which 
the discrepancy itself is resolved without any complicated
unconventional explanation.
Phonon dispersions and changes of nesting vectors in Fermi surfaces 
are clarified to lead to the variety of superlattice structures 
even for the common crystal structures when without CDW,
including orthorhombic $2 \times 2 \times 1$ one for BaTi$_2$As$_2$O,
which has not yet been explained successfully so far, 
being different from tetragonal $\sqrt{2} \times \sqrt{2} \times 1$ 
for BaTi$_2$Sb$_2$O and BaTi$_2$Bi$_2$O.
The electronic structure analysis can naturally explain experimental 
observations about CDW including most latest ones 
without any cramped unconventional mechanisms.
\end{abstract}
\maketitle

\begin{figure}[htbp]
  \centering
  \includegraphics[width=5cm]{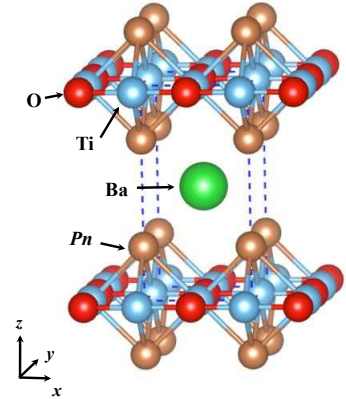}
  \caption{
  Undistorted crystal structures of BaTi$_2${\it Pn}$_2$O ({\it Pn} = As, Sb, and Bi).
  Crystal symmetry is $P4/mmm$ (No.123).
  }
  \label{crystal_structure}
\end{figure}

Layered titanium oxypnictides, 
$A$Ti$_2Pn_2$O [$A =$ Na$_2$, Ba, (SrF)$_2$, 
(SmO)$_2$; $Pn =$ As, Sb, Bi],~{\cite{1997AXE, 2001OZA, 2009LIU, 2010WAN, 2012YAJ, 2012DOA, 2013YAJ1}}
have the common undistorted structure, as shown in Fig.~\ref{crystal_structure},
including Ti$_2$O-plane that leads to
quasi two-dimensional (2D) electronic structures.
Yajima {\it et al.}~\cite{2012YAJ} synthesized BaTi$_2$Sb$_2$O
and reported its superconductivity with the transition
temperature, $T_c = 1.2$ K. 
Doan {\it et al.}~\cite{2012DOA} also synthesized
Ba$_{(1-x)}$Na$_x$Ti$_2$Sb$_2$O individually and reported
its superconductivity with $T_c = 5.5$ K.
Followed by their pioneering works, similar kinds of compounds, 
BaTi$_2$Bi$_2$O, BaTi$_2$(Sb$_{1-x}$Bi$_{x}$)$_2$O, 
BaTi$_2$(Sb$_{1-x}$Sn$_{x}$)$_2$O, Ba$_{1-x}$K$_x$Ti$_2$Sb$_2$O, and
Ba$_{1-x}$Rb$_x$Ti$_2$Sb$_2$O, have been synthesized to get superconductivities, 
achieving the current highest $T_c$ around $6.1$ K.
~\cite{2013YAJ1,2013YAJ2,2013ZHA,2013NAK,2014PAC,2014ROH}
Based on Allen-Dynes formalism~\cite{1957BAR,1965ALL} within DFT,
Subedi suggested a conventional BCS-type superconductivity
mechanism holds in BaTi$_2$Sb$_2$O.~\cite{2013SUB}
This theoretical finding was supported afterward by 
experiments of specific heats, NMR, and $\mu$SR~\cite{2013KIT,2013NOZ,2013ROH}, 
confirming full-gap BCS mechanism with $s$-wave paring for this compound.
Although their $T_c$ values themselves are relatively low compared with
possibly conventional BCS-type mechanism, 
the superconductivity of BaTi$_2$Sb$_2$O and its relatives
attracts special interests in the sense that their nominal 
electronic configurations, Ti$^{3+}$($d^1$), are conjugate with those for 
cuprates superconductors~\cite{1986BED} 
with respect to the electron-hole symmetry.
Quasi 2-dimensional (2D) transports in these systems
also attract the common interest among those for cuprates~\cite{1986BED}
as well as for iron arsenides superconductors~\cite{2008KAM},
leading to the arguments on the similarity of 
superconducting mechanisms.~\cite{2012YAJ,2012DOA}

\vspace{2mm}
The possibilities of the density waves (DW) in these systems
are one of the key concepts, which is common to low-dimensional
transports possible for cuprates and iron arsenides. 
Anomalies in the temperature dependence of resistivity $\rho(T)$ 
and magnetic susceptibility $\chi(T)$ at low temperature are
sometimes observed in these systems, 
getting into an argument over if the anomalies can be attributed to the 
emergence of charge density waves (CDW) or 
spin density waves (SDW).~{\cite{1997AXE, 2001OZA, 2010WAN, 2012YAJ, 2012DOA}}
Several DFT studies applied to BaTi$_2$As$_2$O 
and BaTi$_2$Sb$_2$O~\cite{2012SIN,2013WAN1,2014YU} 
have reported the possibility of SDW, 
while it has not yet been observed experimentally, such as by 
NMR/NQR for Sb$^{121/123}$ and $\mu$SR for
BaTi$_2$As$_2$O and BaTi$_2$Sb$_2$O.~{\cite{2013KIT, 2013NOZ,2013ROH}}
Focusing on CDW, the nesting of Fermi surface matters, 
which is enhanced by 2D nature of Ti$_2$O planes.
While conventional models take a simple picture of 2D transport
with Ti-3$d_{xy}$ orbital only,~{\cite{2012YAJ}} 
Singh~\cite{2012SIN} clarified 
a more complicated 3D shape of Fermi surface in BaTi$_2$Sb$_2$O
by taking into account several Ti-3$d$ orbitals contributions.
A similar shape was also predicted in BaTi$_2$Bi$_2$O, 
theoretically~\cite{2013SUE}.
Indeed, such shapes have been recently observed by 
state-of-the-art Angle-Resolved Photo Emission Spectroscopy (ARPES) 
applied to BaTi$_2$As$_2$O and BaTi$_2$Sb$_2$O 
single crystals.~{\cite{2014Xu, 2016SON}}
Subedi~\cite{2013SUB} worked on BaTi$_2$Sb$_2$O using DFT
and reported the possible lattice instability toward CDW with
$\sqrt{2}\times \sqrt{2} \times 1$ superstructure 
at low temperature.
Experimentally, however, no such superlattice peaks were 
found by neutron and electron diffractions 
for this compound,~\cite{2013NOZ,2014FRA}
reviving the argument over if the anomalies in $\rho(T)$ and 
$\chi(T)$ can be really attributed to the phonon-driven CDW or not.
Though Frandsen {\it et al.}~\cite{2014FRA} observed 
tiny lattice displacements from  tetragonal to orthorhombic in both 
BaTi$_2$As$_2$O and BaTi$_2$Sb$_2$O, 
theoretically predicted $\sqrt{2}\times \sqrt{2} \times 1$ 
superlattice peaks~\cite{2013SUB} could not be found experimentally. 
To account for this, they proposed a bit complicated picture that 
the intra-unit-cell nematic charge order, such as that observed in
cuprates~\cite{2010LAW,2014FUJ}, might be the origin of the
anomalies in $\rho(T)$ and $\chi(T)$.

\vspace{2mm}
Another important topic on this system is the 'two-dome' structure
in the dependence of $T_c(x)$ on the concentration of ionic substitutions, 
$Pn =$ Sb$_x$Bi$_{(1-x)}$ in BaTi$_2$$Pn_2$O.~{\cite{2013YAJ2}}
Similar 'two-dome' structures are known also 
for cuprates~{\cite{1991KOI1,1991KOI2,1991TAM,1992KOI,2004AND} }
and iron arsenides~\cite{2012LIM,2014MAT} superconductors.
The 'two-dome' structure can be regarded as a modification with 
a singularity put on the original single peak dependence. 
The singularity might be attributed to electron or spin
orderings such as DW transition.
Actually, a series of NMR experiments on LaFeAsO$_{1-x}$H$_{x}$~\cite{2013FUJ}
identified the peak as being corresponding to the emergence of
the new SDW phase related
to the second $T_c$ dome. This suggests that another origin of the second dome
be different from that of the first dome.
The 'two-dome' for BaTi$_2$$Pn_2$O, 
on the other hand, has not well been investigated so far.

\vspace{2mm}
As mentioned above, the layered titanium oxypnictides superconductors
show a lot of similar phenomena to cuprates and iron-asenides superconductors. 
The investigation of DW transition in these
oxypnictides could then be one of the most important clues to
understanding of the physics of high-$T_c$ superconducting mechanism. 
In the present study, we investigated the possibility of DW
transition in BaTi$_2Pn_2$O ($Pn =$ As, Sb, Bi)  
using DFT-based phonon analysis.
Because of the common lattice structure when without DW,
we could perform systematic and careful comparisons among 
the three compounds, putting the same computational conditions
to suppress artifacts as less as possible.
As a result, we found a new possibility of
orthorhombic $2 \times 2 \times 1$
superlattice structure for BaTi$_2$As$_2$O, which is
different from the previous prediction by Subedi~\cite{2013SUB} 
for BaTi$_2$Sb$_2$O, tetragonal
$\sqrt{2}\times \sqrt{2} \times 1$.
Our theoretical finding can provide a more natural explanation
for the structural
transition and the weak superlattice peaks observed by Frandsen {\it et al.}
~\cite{2014FRA} in terms of the conventional phonon-driven CDW, 
not by such a complicated mechanism as 
intra-unit-cell nematic charge ordering, as given in
their papers.~\cite{2010LAW,2014FUJ}
The finding might also account for the anomalies of
$\rho(T)$ and $\chi(T)$ being attributed to the lattice instability.
While BaTi$_2$Bi$_2$O does not show anomalies in $\rho(T)$ 
and $\chi(T)$, we found lattice instability possibly inducing 
tetragonal $\sqrt{2} \times \sqrt{2} \times 1$ superlattice structure. 
Such a discrepancy is observed in a LaO$_{0.5}$F$_{0.5}$BiS$ _2$
superconductor as well.~\cite{2012MIZ}
To account for this, Yildirim~{\cite{2013YIL}} argued the possibility 
of an unconventional superconducting mechanism in which the 
inherent lattice instability plays an important role in the Cooper paring.
The similarity of the discrepancy for the present case 
might imply the similar unconventional mechanism for BaTi$_2$Bi$_2$O.

\begin{figure}[htbp]
  \centering
  \includegraphics[width=7cm]{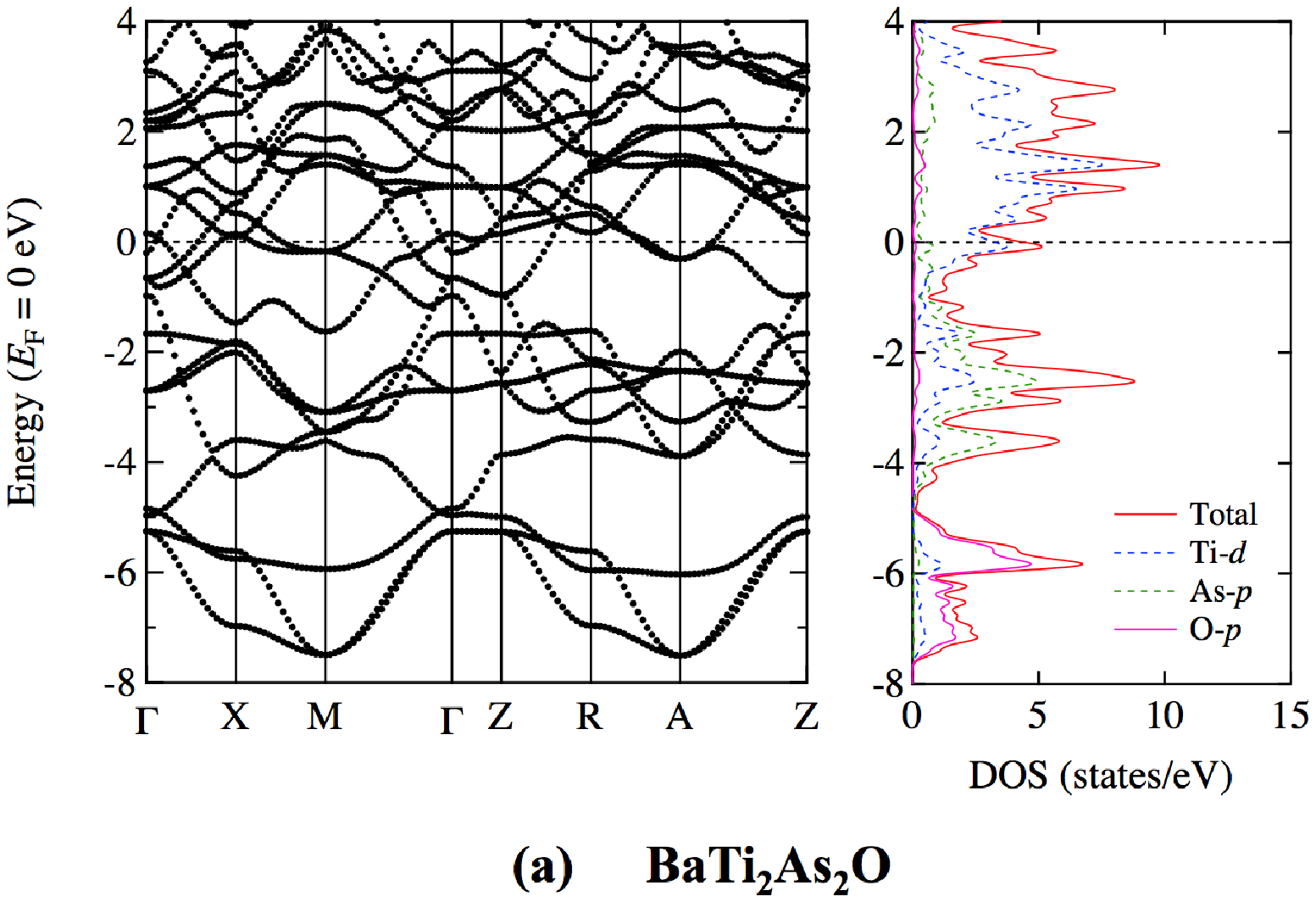}
  \includegraphics[width=7cm]{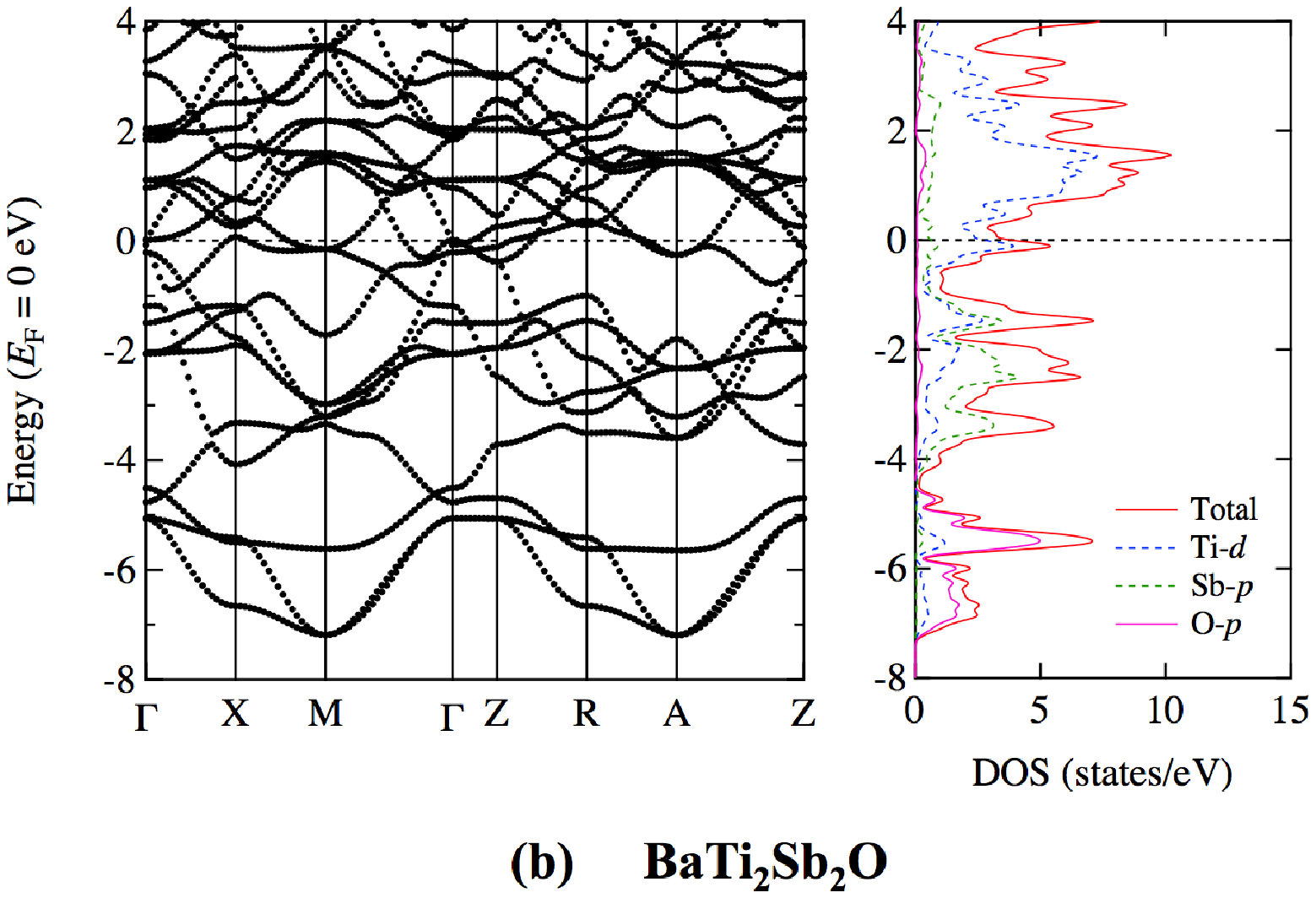}
  \includegraphics[width=7cm]{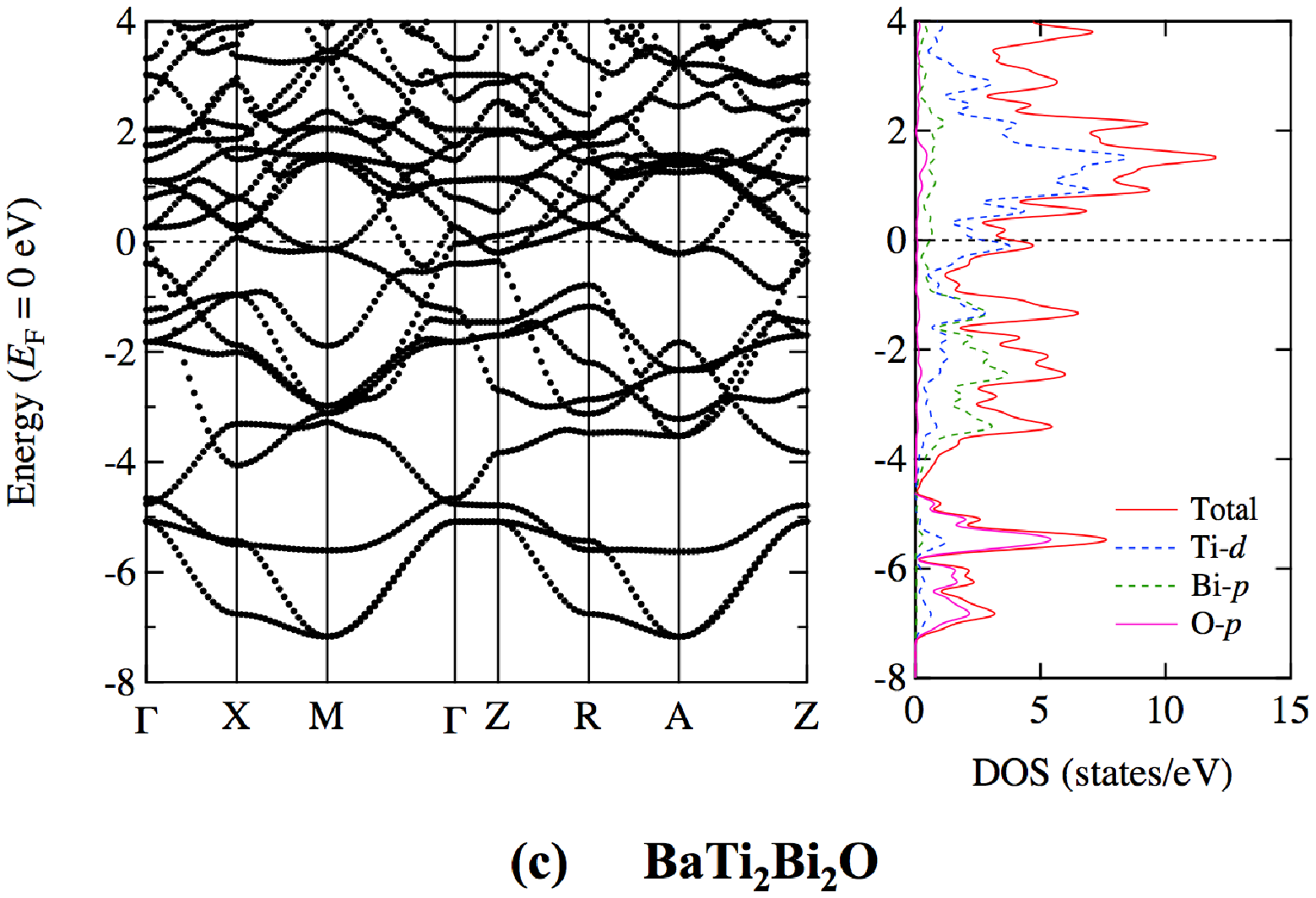}
  \caption{
    Electronic band structures and densities of states
    for BaTi$_2${\it Pn}$_2$O
    ({\it Pn} = (a) As, (b) Sb, and (c) Bi) 
    under $P4/mmm$ symmetry.
    Each Fermi energy is defined as zero.
  }
  \label{btao_bands}
\end{figure}

\section{Results}
\label{sec:results}
\subsection{Lattice instabilities from undistorted structures}
\begin{figure}[t]
  \centering
  \includegraphics[width=8cm]{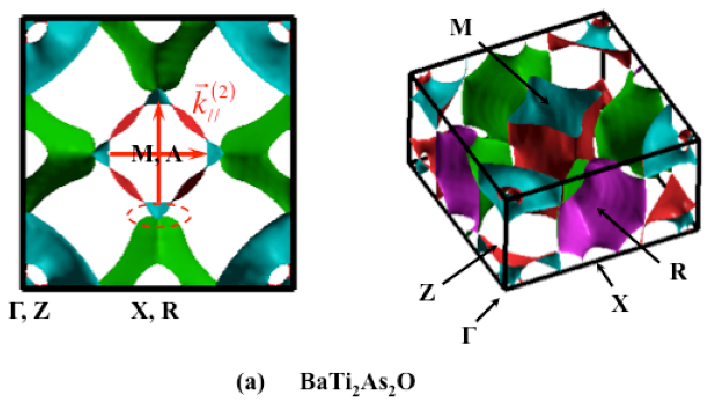}
  \includegraphics[width=8cm]{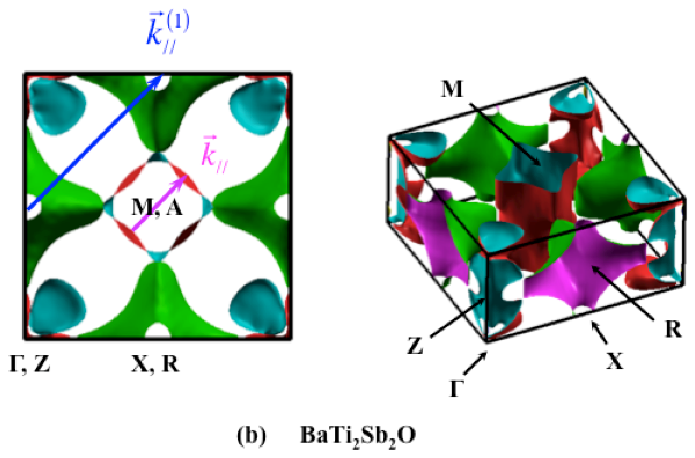}
  \includegraphics[width=8cm]{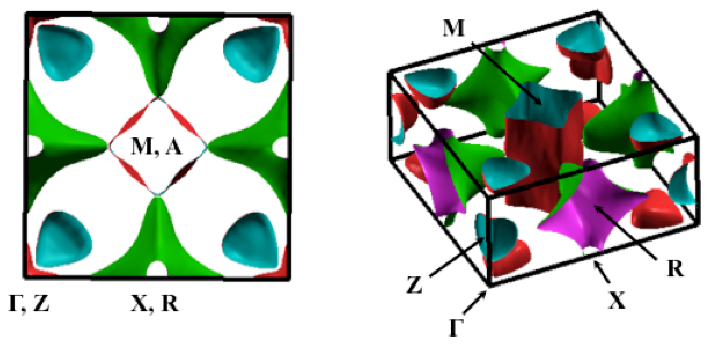}
  \caption{
    Fermi surfaces of BaTi$_2${\it Pn}$_2$O under $P4/mmm$ symmetry 
    ({\it Pn} = (a) As, (b) Sb, and (c) Bi).
    Possible nesting vectors are also depicted 
    (see text for notation).
    Note that $\Gamma$ point is located at the corner,
    as shown in the figure.}
  \label{btao_fermi_surface}
\end{figure}

Electronic band structures and densities of states (DOS)
for undistorted structures are shown in Fig.~\ref{btao_bands}.
Figure~\ref{btao_fermi_surface} highlights
the corresponding Fermi surfaces with possible nesting vectors.
Overall features agree well with previous DFT results 
by Yu {\it et al.}~{\cite{2014YU}}, Singh~{\cite{2012SIN}} and
Suetin {\it et al.}~{\cite{2013SUE}} for each compound,
justifying no specific artifacts due to any choices of
computational conditions.
Note that our optimized geometry parameters are also in good agreements 
with the previous calculations (See Supplementary Note 1).

\begin{figure}[t]
  \centering
  \includegraphics[width=8cm]{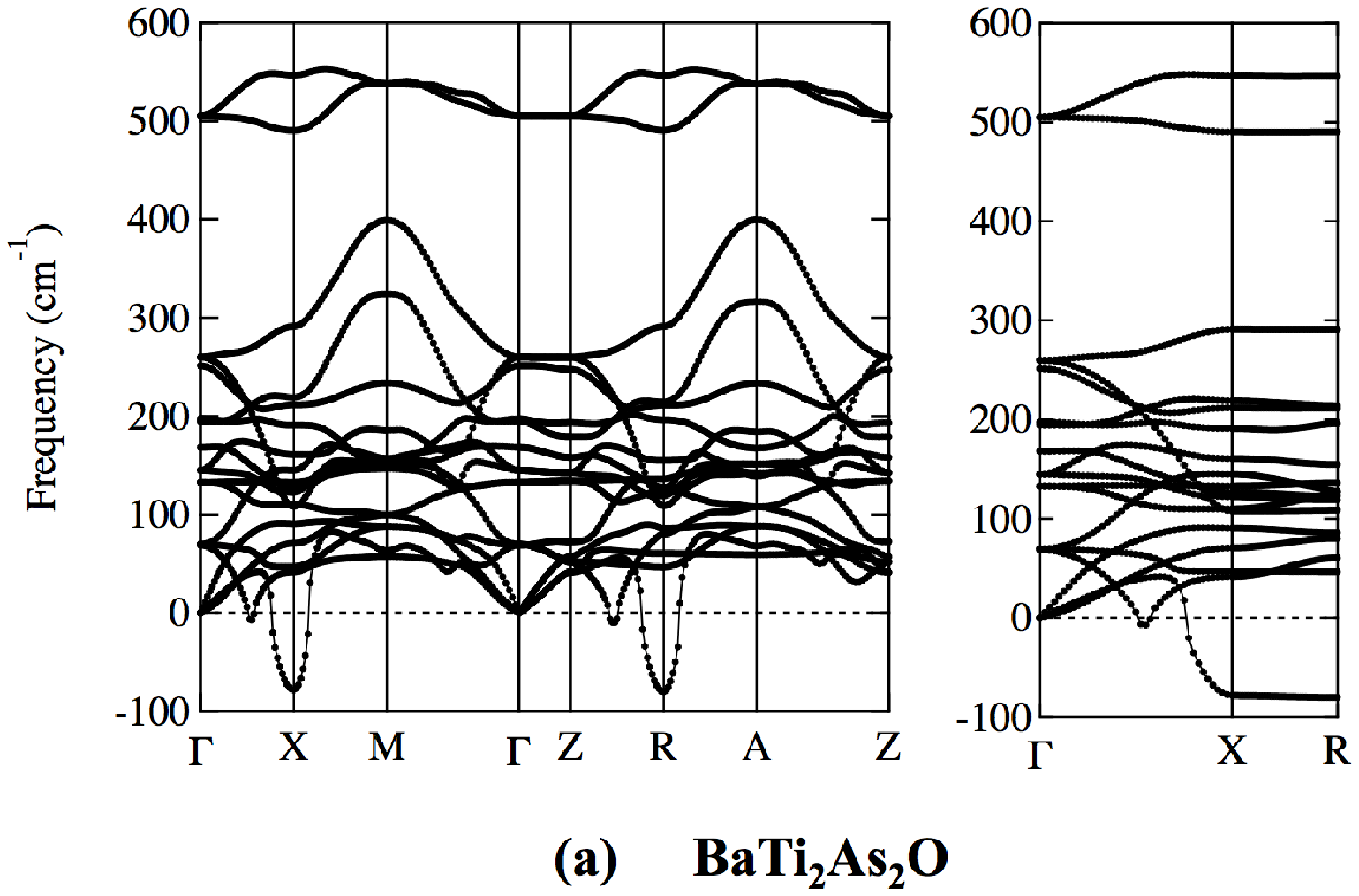}
  \includegraphics[width=8cm]{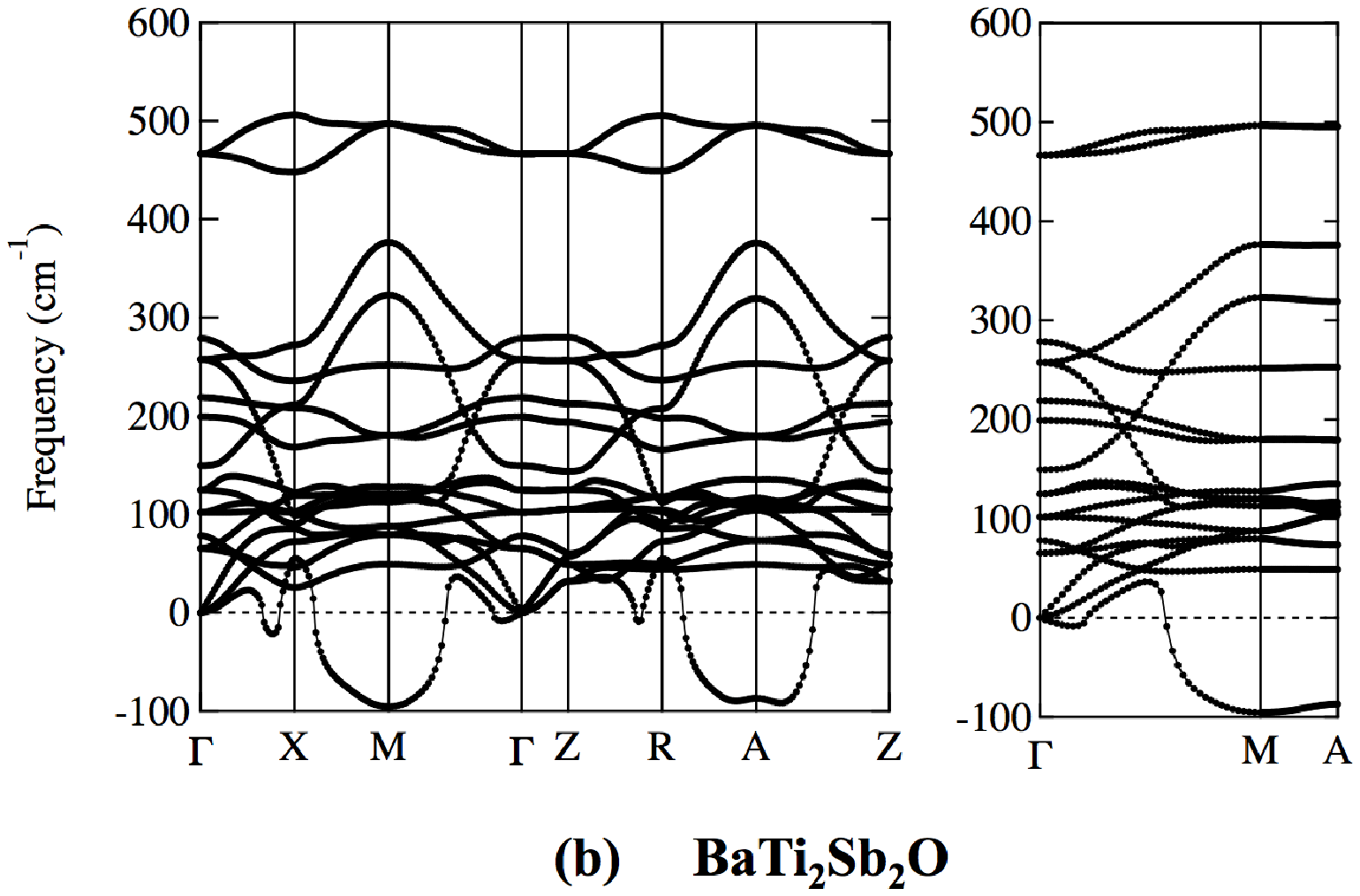}
  \includegraphics[width=8cm]{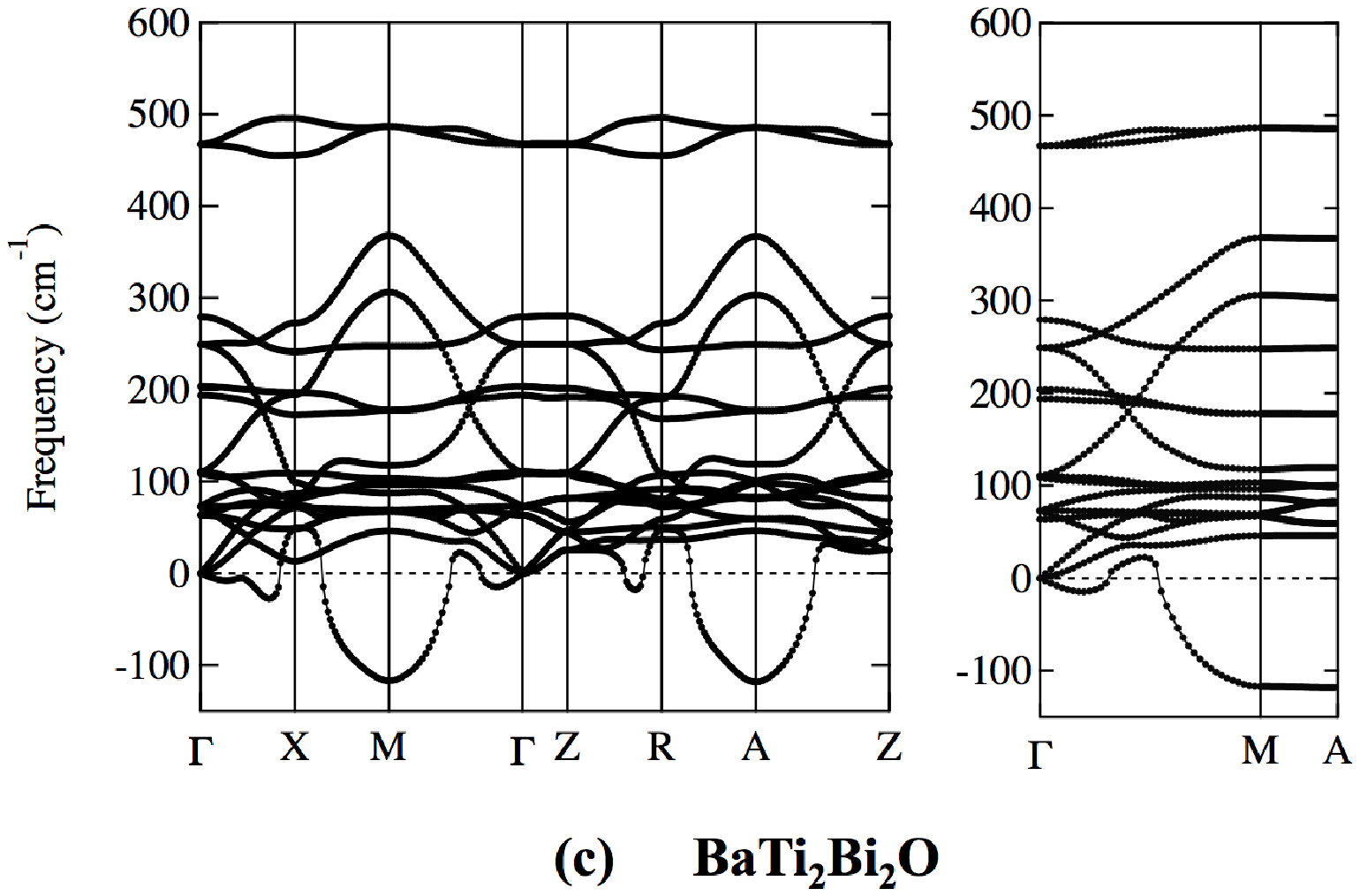}
  \caption{
    Phonon dispersions of BaTi$_2${\it Pn}$_2$O under
    $P4/mmm$ symmetry ({\it Pn} = (a) As, (b) Sb, and (c) Bi).
  }
  \label{btao_ph_bands}
\end{figure}

\begin{figure}[t]
  \centering
  \includegraphics[width=9cm]{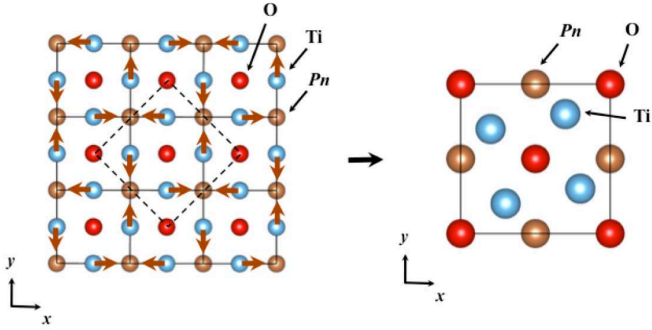}
  \caption{
    Atomic displacements corresponding to the negative (imaginary)
    phonon mode at $M$ $(1/2,1/2,0)$ point, leading to
    $\sqrt{2}\times\sqrt{2}\times 1$ 
    superlattice for BaTi$_2${\it Pn}$_2$O ({\it Pn} = Sb and Bi).
    Dashed lines in the left panel stand
    for the unit cell of the superlattice, drawn in the right panel 
    again with displaced atomic positions after the rotation by 45 degree.
	Symmetry of the superlattice is $P4/mbm$ (No.127).
  }
  \label{btbo_super_lattice}
\end{figure}

\vspace{2mm}
Phonon dispersions for undistorted structures are shown in
Fig.~\ref{btao_ph_bands}.
Despite the common crystal structure, we see that
the phonon instabilities appear around different points,
$M$ and $A$ for $Pn =$ Sb, Bi, while 
$X$ and $R$ for $Pn =$ As
[hereafter, all the {$\vec k$} and {$\vec q$} points are labeled according to 
Brillouin-zone database on the {\it Bilbao Crystallographic 
Server}~{\cite{2014ARO}}].
The former instabilities (for $Pn =$ Sb, Bi) are
consistent with the previous calculations by
Subedi~\cite{2013SUB} on $Pn$ = Sb.
The latter instability (for $Pn =$ As), 
on the other hand, has been never reported before, so we carefully
examined to confirm that the result does not depend on
the choice of pseudo potentials. 
For $Pn =$ Sb and Bi, the instability occurs around $M$ and $A$,
corresponding to $(q_x,q_y)=(1/2,1/2)$
[hereafter a unit of $\vec q$ is $2\pi/a$.].
For $q_z$ direction, there is no specific dependence, as shown in
the dispersion along $M$ to $A$ in the right panel of
Fig.~\ref{btao_ph_bands} (b) and (c).
From phonon pDOS (partical DOS), we can identify which
vibration modes lead to the instability toward the superlattice.
It is found from phonon pDOS that the negative (imaginary) 
frequencies mainly come from Ti 'in-plane' (within $xy$ plane) 
vibrations (See Supplementary Note 2).
We therefore concentrate on the representative case with $M$,
$(q_x,q_y,q_z)=(1/2,1/2,0)$, corresponding to
$\sqrt{2} \times \sqrt{2} \times 1$ superlattice structure.
By analyzing the dynamical matrices, we can further identify
the superlattice structure shown in Fig.~\ref{btbo_super_lattice}.
Note that this is the same structure as that predicted previously by
Subedi~\cite{2013SUB} for $Pn =$ Sb.

\begin{figure}[t]
  \centering
  \includegraphics[width=8cm]{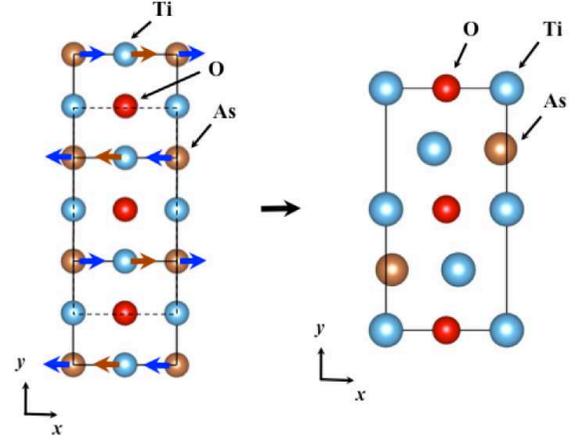}
  \caption{
    Atomic displacements corresponding to the negative (imaginary)
    phonon mode at $X$ $(0,1/2,0)$ point, leading to $1 \times 2 \times 1$ 
    superlattice for BaTi$_2$As$_2$O. Dashed lines in the left panel stand
    for the unit cell of the superlattice, drawn in the right panel 
    again with displaced atomic positions.
	Symmetry of the superlattice is $Pbmm$ (No.51).
    }
  \label{btao_super_lattice}
\end{figure}

\vspace{2mm}
The same scheme is applied to $Pn =$ As with the instability around
$X$ and $R$, corresponding to $(q_x,q_y)=(0,1/2)$: 
For $q_z$ direction, there is no specific dependence, as shown in
the dispersion along $X$ to $R$ in the right panel of
Fig.~\ref{btao_ph_bands} (a).
From phonon pDOS, we can identify that
the vibrations for the instability come from
Ti and As 'in-plane' vibrations (See Supplementary Note 2).
We therefore take a representative mode at $X$, $(q_x,q_y,q_z)=(0,1/2,0)$.
  By analyzing the dynamical matrices, we get the superlattice structure
shown in Fig.~\ref{btao_super_lattice},
$1 \times 2 \times 1$ superlattice.

\subsection{Superlattice structures and $T_c$}
\begin{figure*}[t]
  \centering
  \includegraphics[width=5.5cm]{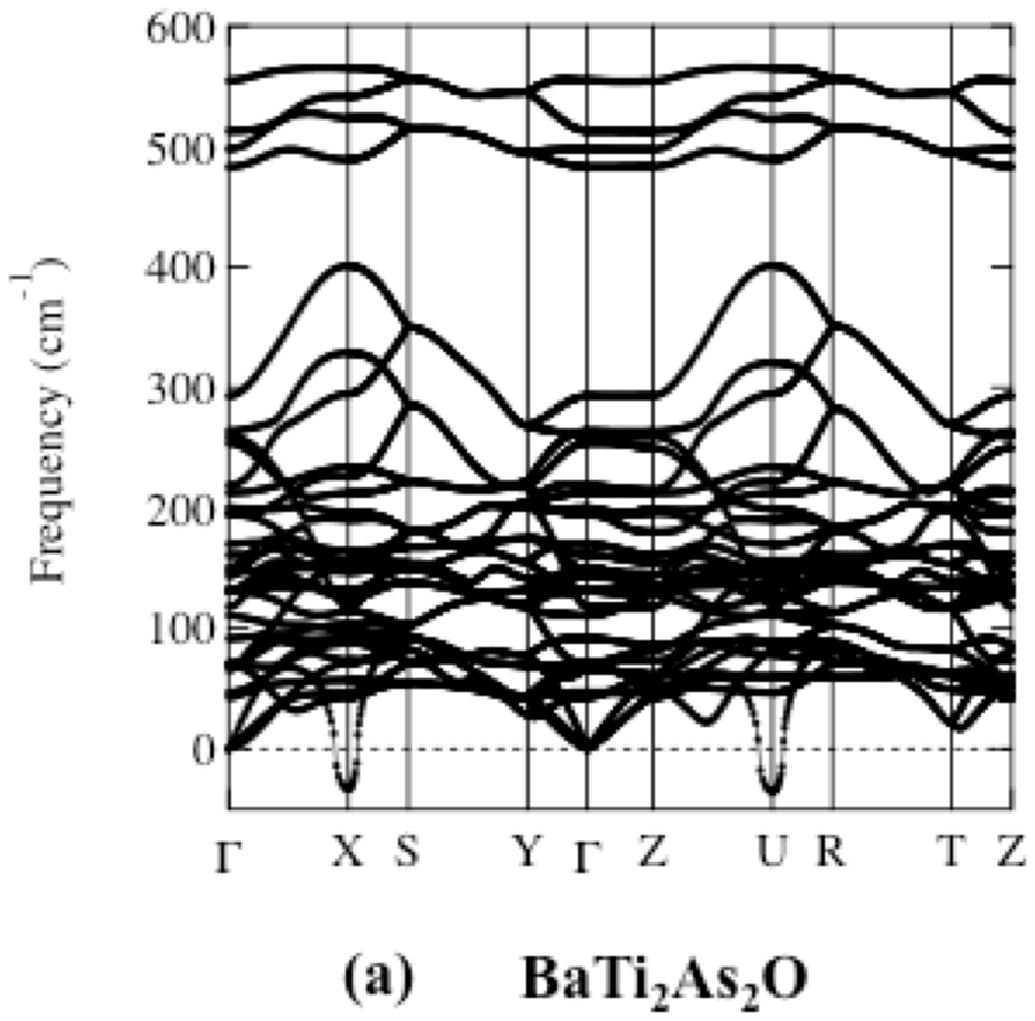}
  \includegraphics[width=5.5cm]{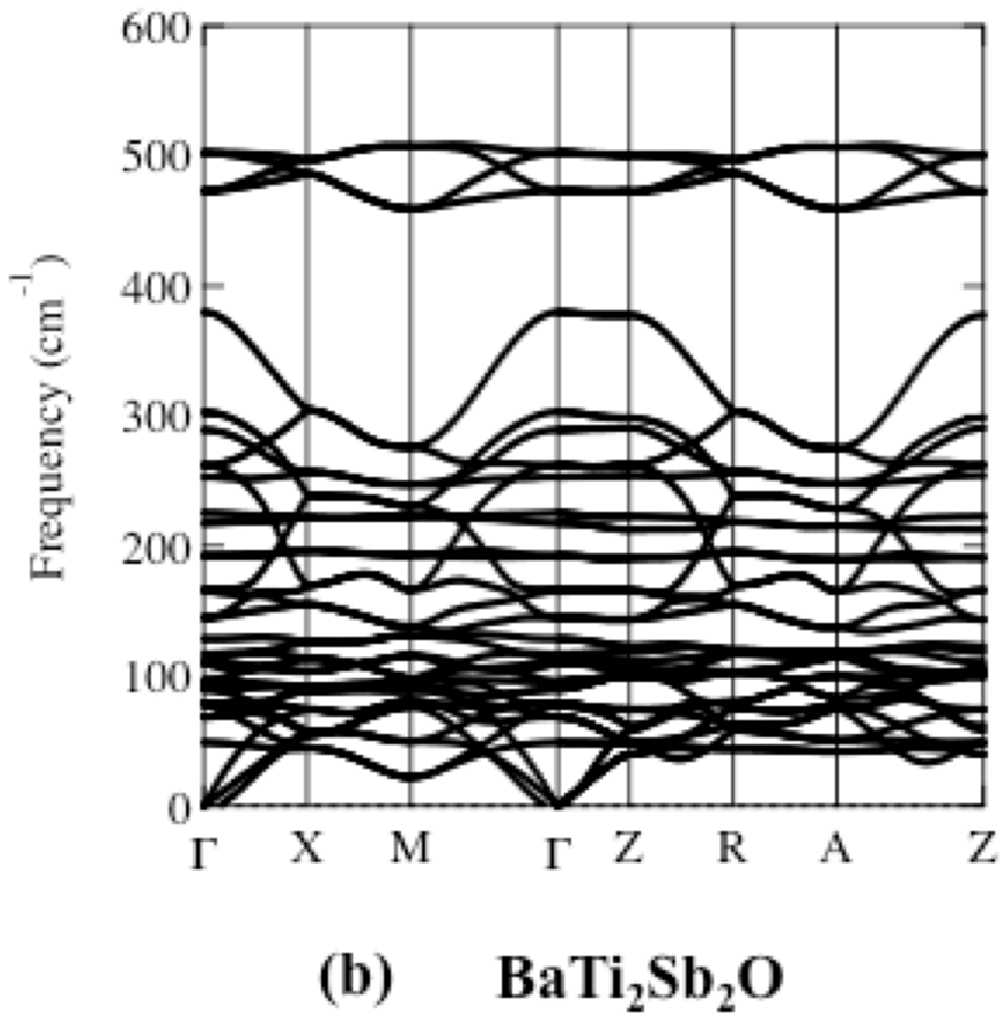}
  \includegraphics[width=5.5cm]{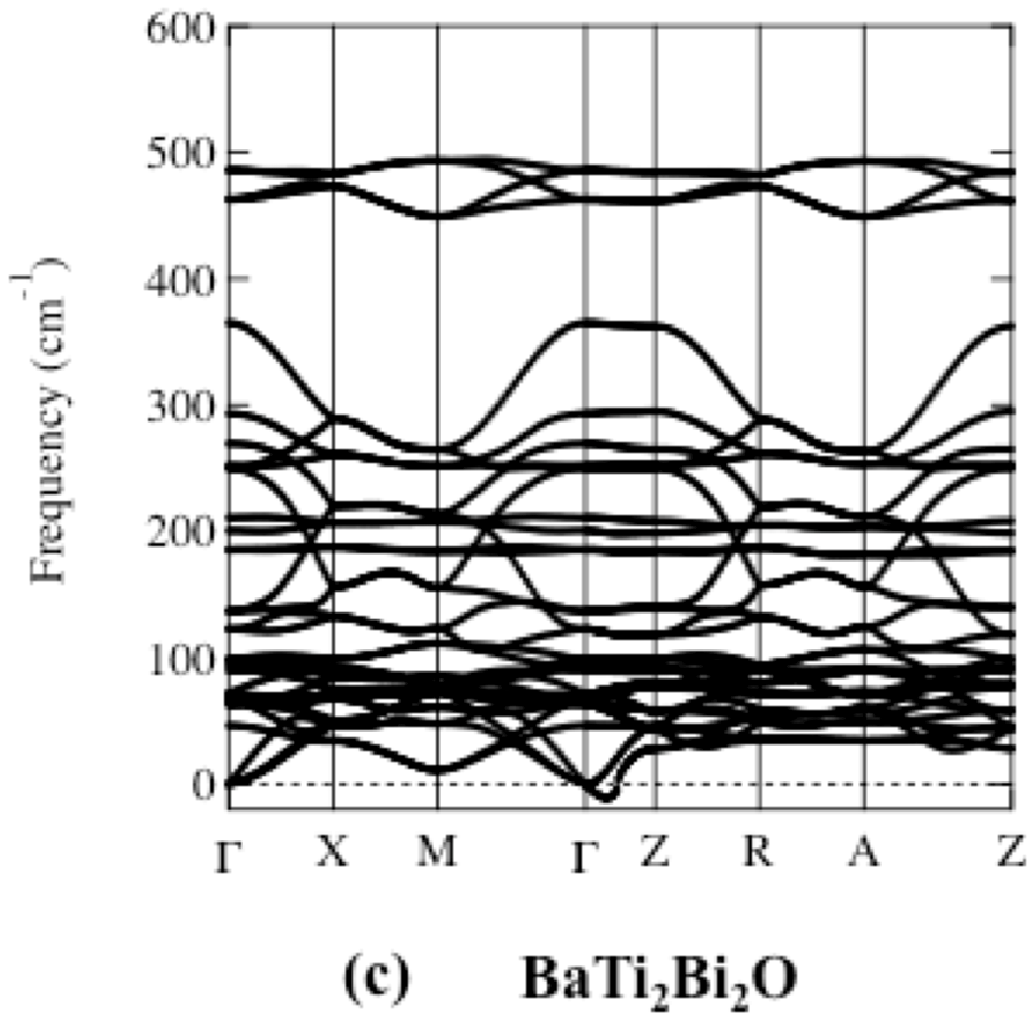}
  \caption{
   Phonon dispersions of superlattice structure for 
   BaTi$_2${\it Pn}$_2$O ({\it Pn} = (a) As-$1 \times 2 \times 1$, 
   (b) Sb-$\sqrt{2}\times\sqrt{2}\times 1$, and (c) Bi-$\sqrt{2}\times\sqrt{2}\times 1$).
   }
  \label{ph_disp_super_lattice}
\end{figure*}

\begin{figure}[t]
  \centering
  \includegraphics[width=8cm]{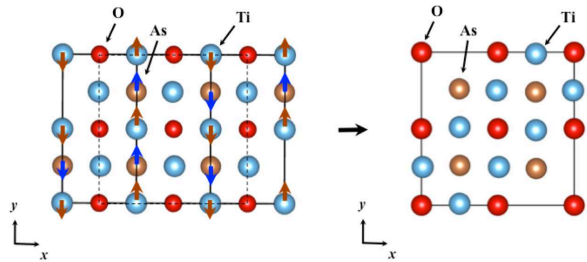}
  \caption{
    Atomic displacements for $1 \times 2 \times 1$ BaTi$_2$As$_2$O 
	corresponding to the negative (imaginary) phonon mode at 
	$X$ $(1/2,0,0)$ point, leading to orthorhombic $2 \times 2 \times 1$
        superlattice. 
	Dashed lines in the left panel stand
    for the unit cell of the superlattice, drawn in the right panel 
    again with displaced atomic positions.
	Symmetry of the $2 \times 2 \times 1$ superlattice is $Pbam$ (No.55).
    }
  \label{btao_2_2_1_super_lattice}
\end{figure}

The negative modes appearing in Fig.~\ref{btao_ph_bands} are expected
to disappear when we further relax the lattices 
along the negative modes to get superlattices. 
Resultant phonon dispersions are shown in 
Fig.~\ref{ph_disp_super_lattice}. 
For $\sqrt{2}\times \sqrt{2} \times 1$ superlattices 
of Bi and Sb, the negative modes have disappeared 
assuring the superlattice as the final stable structures 
[the negative mode seen around $\Gamma$ for Bi 
is due to the well-known artifact coming from the discreteness of 
Fast Fourier Transform (FFT) grid].
For As, on the other hand, the negative modes at 
$X$ and $U$ point still remain.
Taking the representative $X$ point, this implies the further 
superlattice transition toward $2 \times 2 \times 1$.
The phonon pDOS analysis shows that the superlattice 
deformation still comes from 'in-plane' vibrations 
of Ti and As, actually leading to the superlattice structure as shown in 
Fig.~\ref{btao_2_2_1_super_lattice}.
The geometry optimization along this deformation gives 
lattice parameters, $a = 8.122$ $\AA$, $b = 8.108$ $\AA$, 
and $c = 7.401$ $\AA$ (See Supplementary Note 4).
The orthorhombicity parameter is $\eta = 2 \times (a-b)/(a+b) = 0.171 \%$. 
To finalize the verification, we ought to examine if 
the negative modes surely disappear in the 
$2 \times 2 \times 1$ superlattice. 
It was, however, intractable 
because four times enlarged unit cell requires $4^3$ times more computational cost
and lower symmetry makes more demands for $k$- and $q$-mesh samplings
over larger reciprocal space.

\vspace{2mm}
Superconducting transition temperatures, $T_c$, 
are estimated
using Allen-Dynes formula~\cite{1965ALL}, 
as shown in Table VI in Supplementary Information. 
The parameters used in the formula are also 
tabulated. 
Even with imaginary frequencies, 
$T_c$ can be estimated just by ignoring the 
contributions~\cite{2013SUB}, 
being the case for $1 \times 1 \times 1$. 
For $\sqrt{2} \times \sqrt{2} \times 1$ where 
all the negative modes disappear, 
the estimation gets to be more plausible. 
There seems, however, little change in the 
estimation from that in $1 \times 1 \times 1$. 
The present estimations, $T_c = 2.30~(2.45)$ K 
for $Pn =$ Sb (Bi), 
are consistent with experiments reporting 
$T_c = 1.2~(4.6)$ K for $Pn =$ Sb (Bi)~{\cite{2012YAJ,2013YAJ1}}, 
as well as that by previous DFT study, 
$T_c = 2.7$ K for $Pn =$ Sb~{\cite{2013SUB}}.
For $1 \times 2 \times 1$ ($Pn$ = As), 
the estimation was made still under the 
existence of negative modes, and 
the more plausible estimation for 
 $2 \times 2 \times 1$ was intractable 
as mentioned above.
The estimations, $T_c$ = 6.93 K ($1\times 1\times 1$)
and 8.31 K ($1\times 2\times 1$) seem incompatible
with experiments of BaTi$_2$As$_2$O 
exhibiting no superconductivity so far.~\cite{2010WAN, 2013YAJ2}
We might expect much lower $T_c$ obtained for 
the $2\times 2\times 1$ superlattice structure.

\section{Discussions}
\label{sec:discussions}
\subsection{Comparison with experiments}
The orthorhombic $2\times 2\times 1$ superlattice structure,
obtained here for BaTi$_2$As$_2$O, would attract interests 
in connection with experimental observations.
Frandsen {\it et al.} actually observed a lattice 
structural transition from tetragonal $P4$/$mmm$ to orthorhombic $Pmmm$ by
neutron diffractions. They also reported very weak superlattice peaks corresponding to 
$q_{Hx}$ = (1/2, 0, 0) or $q_{Hy}$ = (0, 1/2, 0) which disappears
at higher temperatures
by electron diffractions.~{\cite{2014FRA}}
Since BaTi$_2$As$_2$O has the same $P4$/$mmm$ parent structure
as BaTi$_2$Sb$_2$O, 
they might expect a tetragonal
$\sqrt{2}\times \sqrt{2} \times 1$ superlattice 
according to the previous DFT prediction by
Subedi~{\cite{2013SUB}} for $Pn =$ Sb.
The lack of the expected $\sqrt{2}\times \sqrt{2} \times 1$
superlattice and the observation of the structural transition
were explained by rather complicated mechanism such as intra-unit-cell
nematic charge ordering, as an analogue of that
in cupurates.~\cite{2010LAW,2014FUJ}
The observed weak peaks of $q_{Hx}$ or $q_{Hy}$ was then regarded as 
being non-intrinsic, attributed to poly-crystalline grain
boundary effects.
Our finding of orthorhombic $2\times 2\times 1$ superlattice
here can instead account for the observations,
as intrinsic, more naturally in terms of 
the lattice instability due to conventional phonon-driven CDW.
Note that our optimized geometry gives
the orthorhombicity parameter,
  $\eta = 2 \times (a-b)/(a+b) = 0.171 \%$,
which is comparable
with the experimental value $\eta = 0.22 \%$~{\cite{2014FRA}}.
Though $q_{Hx}$ or $q_{Hy}$ is naturally explained, 
the present result also leads to the emergence of 
$q_{Hxy}$ = (1/2, 1/2, 0), which is not explicitly 
reported in the work by Frandsen {\it et al.}~{\cite{2014FRA}}
Looking at their TEM photo in the paper,~{\cite{2014FRA}}
it is actually quite difficult to distinguish 
the $q_{Hxy}$ peak from much brighter spots in the immediate vicinity.
Polycrystalline sample qualities and weak intensities of the peak 
might also matter. 
We expect that further careful investigation would 
find $q_{Hxy}$ peak corresponding to the present 
$2\times 2 \times 1$ superlattice structure. 

\vspace{2mm}
Unlike $Pn =$ As, the other two compounds are predicted to
have $\sqrt{2} \times \sqrt{2} \times 1$ superlattices
in our calculations, which is consistent with the preceding 
work by Subedi~\cite{2013SUB} for $Pn =$ Sb.
Then a question arises asking why only $Pn =$ As takes the
different superlattice structure.
This can be explained by the nesting of 
Fermi surfaces, shown in Fig.~\ref{btao_fermi_surface}.
Because of the cylinderical shape, every compound has a nesting 
vector $\vec k_{//}=(k_x,k_y)=(1/4,1/4)$ 
[hereafter a unit of $\vec k$ is $2\pi/a$]. 
around $M$ and $A$ points, 
as previously pointed out 
by Yu {\it et al.}~{\cite{2014YU}} for $Pn =$ As .
Another possible nesting around $X$ and $R$ is described 
by the vector $\vec k_{//}^{(1)}=(1/2,1/2)$, corresponding
to the negative phonons we get around $M$ and $A$ in phonon 
Brillouin zones for $Pn =$ Sb, Bi. 
This nesting has already been pointed out by Singh~{\cite{2012SIN}}
and Subedi~{\cite{2013SUB}} for $Pn =$ Sb.
Looking carefully at the Fermi surface of $Pn =$ As, we see 
the flattening of the cusp around $X$ and $R$ points (as shown 
by a dashed oval in Fig.\ref{btao_fermi_surface} (a)), 
leading to new nesting vectors, $\vec k_{//}^{(2)}=(0,1/2)$
and $(1/2,0)$. They correspond to the negative phonons 
around $X$ and $R$ appearing only for $Pn =$ As. 
We note that $\vec k_{//}^{(2)}$ have already 
been mentioned by Yu {\it et al.}~{\cite{2014YU}} for $Pn =$ As, but 
its relation to the phonon instability has not been discussed so far.
Possible reasons why the flattening occurs only for As
are discussed in the next section.
Since all the above stories can be made only within the electronic
Fermi surfaces, one might consider the phonon evaluations not necessarily
required.
We note, however, that there are several 2D chalcogenide systems 
where CDW cannot be explained only by the electronic 
Fermi surfaces, but accounted for when the phonon dispersions 
are evaluated.~\cite{2006JOH,2008JOH,2009CAL,2011CAL,2015ZHU}
{We discuss this in the later section.}
It is further interesting if CDW superlattice transitions 
predicted here could be related to the anomalies of 
$\rho(T)$ and $\chi(T)$ at low temperature. 

\vspace{2mm}
The $\sqrt{2} \times \sqrt{2} \times 1$ superlattice structure
for BaTi$_2$Sb$_2$O
is predicted not only by the present work but also by Subedi~{\cite{2013SUB}}.
However, such a superlattice has not yet been observed
experimentally so far by any diffraction experiments such as
neutron and electron diffractions.~\cite{2013NOZ,2014FRA}
Frandsen {\it et al.} observed a subtle 
structural distortion from $P4/mmm$ to $Pmmm$ by neutron diffraction 
measurement at low temperature,~{\cite{2014FRA}} which is not consistent with
the prediction.
Note that all the above diffraction experiments are applied to
polycrystalline samples.
Interestingly, Song {\it et al.} has very recently reported 
the existence of $(1/2,1/2)$ nesting vector 
by ARPES and STM measurements applied to
high-quality single crystals, being consistent with the theoretical
predictions.~\cite{2016SON}
The superlattice structure is hence expected to be detected
by further diffraction measurements.
The theoretically estimated distortion here is quite small,
$0.14$ \AA{} (See Supplementary Note 4),
being in agreement with the previous calculation by Subedi,~{\cite{2013SUB}}
so careful detections would be required for experiments.
As described in the previous section,
the estimated $T_c$ is almost consistent with experiments, supporting 
that the compound is a conventional BCS-type superconductor, 
being consistent with the previous 
conclusion by Subedi.~{\cite{2013SUB}}

\vspace{2mm}
In contrast to the other two compounds, 
there are little experiments on BaTi$_2$Bi$_2$O because of 
the difficulty of sample preparation mainly due to the 
significant instability under the air and moisture.~{\cite{2013YAJ1}}
As far as we have known, there is no 
previous research on phonon dispersions on this system.
The anomalies of $\rho(T)$ and $\chi(T)$ disappear 
with increasing $T_c$ when Sb is gradually substituted by Bi, 
as reported experimentally.~{\cite{2013YAJ1}}
Though there is no direct evidence by diffraction 
experiments, it seems, then, the present consensus 
on this compound that there is no instability 
toward CDW, which would be contradicting to our 
result here.
The spin polarization, not taken into 
account here, may be one of the possibilities to modify 
the nesting, for instance via the spin-orbit coupling, 
accounting for this discrepancy, but 
it is reported, at least for DOS, 
the effect matters little.~\cite{2012SIN}
A similar discrepancy between negative phonon
predictions~\cite{2013YIL, 2013WAN2} 
and unobserved structural instability is known for 
a superconductor, LaO$_{0.5}$F$_{0.5}$BiS$_2$.~{\cite{2012MIZ}}
A large phonon instability toward a static CDW was estimated
theoretically,~{\cite{2013YIL}}
while no anomaly in $\rho(T)$ and $\chi(T)$ has been observed
experimentally.~{\cite{2012MIZ, 2013LEE}}
Yildirim~{\cite{2013YIL}} then argued the possibility of an unconventional 
superconducting mechanism in which inherent lattice 
instabilities have an important role on the Cooper paring
in this compound.
A more recent neutron diffraction experiment~{\cite{2015ATH}}
reported that the local distortion of the atomic position of S 
around Bi is detected under $T_c$, attracting an attention 
in connection with the unconventional mechanism. 
In our case of BaTi$_2$Bi$_2$O, the stabilization energy is 
evaluated around $23.3$ meV/UnitCell, being much larger than that of 
LaO$_{0.5}$F$_{0.5}$BiS$_2$ ($\sim$10 meV/UnitCell).~{\cite{2013YIL}}
In terms of the magnitude of the displacement, it is 
$0.16$ \AA{} for BaTi$_2$Bi$_2$O (See Supplementary Note 4), 
which is the same as $0.16$ \AA{} for
LaO$_{0.5}$F$_{0.5}$BiS$_2$.~\cite{2013YIL}
In addition, the $\sqrt{2} \times \sqrt{2} \times 1$ 
superlattice structure obtained by analyzing dynamical matrices for BaTi$_2$Bi$_2$O 
does not show any negative phonon frequency (Fig.~\ref{ph_disp_super_lattice} (c)).
Therefore, the lattice instability is also expected to be static. 
The similarity might imply the similar unconventional 
mechanism also for BaTi$_2$Bi$_2$O. 
If it were so, the substitution of Sb by Bi would introduce 
the unconventionality to the conventional BCS of
BaTi$_2$Sb$_2$O.~{\cite{2013KIT,2013NOZ,2013ROH}}
The introduction might account for the two-dome structure appeared 
by the substitution.~{\cite{2013YAJ2}}

\subsection{Possible mechanism for variety of superlattices}
A natural question consequently arising would be 
asking why the new nesting vectors $\vec k^{(2)}_{//}$ 
appear only for $Pn$ = As. 
The vectors are caused by the flattening of the 'nose' 
of Fermi surfaces directing toward the central cylinder 
from four equivalent outsides.
Interestingly, the similar flattened 'nose' was actually 
reported in the paper by Singh {\it et al.} \cite{2012SIN} 
(in their Fig. 7), shown as the 'Fermi surfaces' below 
$E_F$ by 0.1~eV for $Pn$ = Sb. 
Looking at our Fig.~\ref{btao_bands}, 
we observe that the Fermi level seems to be 
approaching down toward the DOS peak as $Pn$ 
changes from Bi and Sb to As. 
This can be regarded as if $E_F$ effectively 
behaves like the 'sea-level down' $Pn$ = As 
with a fixed landscape of $Pn$ = Sb.
The Fermi surface of $Pn$ = As would therefore 
correspond to that of $Pn$ = Sb with negative 
energy shift, as shown in 'Fig. 7 by 
Singh {\it et al.}'. \cite{2012SIN}
The 'fixed landscape', namely the 'rigid band picture' 
near to Fermi level, can be justified to some extent 
because they are mainly composed of Ti-$d$ orbital 
contributions as shown in Fig.~\ref{btao_bands}.
The reason why the 'sea-level' gets down when 
$Pn$ is substituted into As can roughly be 
accounted as follows: 
As a rough estimation of how $Pn$ affects to shift 
$E_F$, we can start with its 'HOMO' level of the
isolated atom 
[HOMO stands for 'Highest Occupied Molecular Orbital'
though the orbitals in the present context is not molecular
but isolated atomic orbital. We use 'HOMO' rather than HOAO
just because the latter is not so commonly spread abbreviation.
We expect this doesn't matter so much
even it is used for isolated atom.].
Namely, a negatively deeper 'HOMO' would contribute 
to attract Ti-electrons more strongly and make $E_F$ lower.
Noticing the deeper 'HOMO' corresponds to 
the larger ionic potential, we expect that 
the lighter element (As) has deeper 'HOMO' because 
the potential is enhanced by the less 
screening of the nucleus attractions 
by fewer inner electrons.~\cite{2004HON} 
The deeper 'HOMO' also corresponds to the 
larger electronegativity, which is actually 
2.18 for As while 2.05 (2.02) for Sb (Bi) by Pauling scale. 
Similar negativities for Sb and Bi 
can account for the common nesting vectors 
of these compounds, being different from 
that of As.
Summarizing the above, the negatively deeper 
'HOMO' level of As can attract Ti-electrons more strongly and 
effectively push $E_F$ down when it forms pnictides, and then 
the Fermi surface changes to get flattened 'nose' 
as depicted in 'Fig. 7 of Singh {\it et al.}~\cite{2012SIN}'.

\vspace{2mm}
Though we could not make clear explanations here, 
we must note that 
the nesting vector cannot solely account for 
the superlattice instabilities even in the present case.
In addition to $\vec k^{(2)}_{//}$ for $Pn = $ As, 
$\vec k^{(1)}_{//}$ and $\vec k_{//}$ may be regarded
as possible nesting vectors.
The Kohn anomalies corresponding to 
$\vec k^{(1)}_{//}$ and $\vec k_{//}$ are, however, 
not present in the phonon dispersion.
This fact might be related to
recent intensive discussions about the Kohn anomaly~\cite{1959KOH}: 
Some studies \cite{2006JOH,2008JOH,2009CAL,2011CAL,2015ZHU} insist that
the imaginary phonons in quasi 2D systems
are dominated not mainly by 
the nesting of electronic structures but 
rather primarily by the wave vector dependence of 
the electron-phonon coupling, $g(\vec q)$. 
There exists, however, such quasi 2D systems~\cite{2015KIM} 
where their superlattice instabilities can 
clearly be explained by the nesting vectors. 
To investigate if our system corresponds to 
which case or that lying in between,
it is quite intriguing to analyze $g(\vec q)$
for $Pn$ = As, but unfortunately we cannot 
perform any of such phonon calculations under the 
perfect disappearance of imaginary frequencies 
because of too costly calculations for 
the $2\times 2\times 1$ superlattice.

\section{Methods}
\label{sec:methods}
All the calculations were done within DFT 
using GGA-PBE exchange-correlation functionals~\cite{1996PER}, 
implemented in Quantum Espresso package.~\cite{2009PAO}
After carefully examining the artifacts due to
the choice of pseudo potentials (PP),
we provide here the final results mainly 
obtained by the PAW~\cite{1994BLO}
framework of the valence/core separation of electrons. 
The implementation of PAW adopted here 
takes into account the relativistic effects 
within the extent of the scalar-relativistic theory
upon a careful comparison with all-electron calculations
by Wien2k.~{\cite{2014KUC}}
We restricted ourselves to spin unpolarized calculations,
anticipating that the spin polarization affects little as 
supported by several experiments.~\cite{2013KIT,2013NOZ,2013ROH}
Lattice instabilities are detected by the negative (imaginary) 
phonon dispersions evaluated for undistorted lattice structures.
Taking each of the negative phonon modes, the structural relaxations along
the mode are evaluated by the BFGS optimization scheme with the 
structural symmetries fixed to $Pbmm$ ($1 \times 2 \times 1$) and 
$Pbam$ ($2 \times 2 \times 1$) for BaTi$_2$As$_2$O, $P4/mbm$ ($\sqrt{2} \times \sqrt{2} \times 1$)
for BaTi$_2$Sb$_2$O and BaTi$_2$Bi$_2$O. 
For phonon calculations, we used the 
linear response theory implemented in Quantum Espresso package.~{\cite{2001BAR}}
Crystal structures and Fermi surfaces are depicted by using 
VESTA~\cite{2011MOM} and XCrySDen~\cite{1999KOK}, respectively.

\vspace{2mm}
To deal with the three compounds systematically, we took the same conditions 
for plane-wave cutoff energies ($E_{\rm cut}$), $k$-meshes, 
and smearing parameters.
The most strict condition among the compounds is taken 
to achieve the convergence within $\pm 1.0$ mRy in the
ground state energies of undistorted (superlattice) 
systems, resulting in 
$E_{\rm cut}^{(\rm WF)} = 90~ (100)$ Ry 
for wavefunction and
$E_{\rm cut}^{(\rho)} = 800~ (800)$ Ry 
for charge densities.
For $T_c$ evaluation, we adopted unshifted $k$-meshes
centered at $\Gamma$-point.
Denser $k$-meshes should be taken for electron-phonon calculations 
because of the double-delta integrations.~{\cite{2008WIE}}
For undistorted systems, ($8 \times 8 \times 4$) $k$-meshes were used for the
Brillouin-zone integration. 
Phonon dispersions were calculated on ($8 \times 8 \times 4$) $q$-meshes.
Denser $k$-meshes, ($24 \times 24 \times 12$), 
were used for the double-delta integrations in electron-phonon calculations. 
For distorted BaTi$_2$As$_2$O superlattices, ($8 \times 4 \times 4$) and ($4 \times 4 \times 4$) $k$-meshes
were used for $1 \times 2 \times 1$ and $2 \times 2 \times 1$ superlattices, respectively.
Phonon dispersions were calculated on ($8 \times 4 \times 4$) and ($4 \times 4 \times 4$) $q$-meshes.
Denser $k$-meshes, ($24 \times 12 \times 12$), were used for the double-delta integrations in 
electron-phonon calculations of the $1 \times 2 \times 1$ superlattice. 
For distorted BaTi$_2$Sb$_2$O and BaTi$_2$Bi$_2$O superlattices, ($6 \times 6 \times 6$) $k$-meshes 
are used. Phonon dispersions were calculated on ($6 \times 6 \times 6$) $q$-meshes.
Denser $k$-meshes, ($18 \times 18 \times 18$), were used for the double-delta integrations in 
electron-phonon calculations.
The Marzari-Vanderbilt cold smearing scheme~\cite{1999MAR} with
a broadening width of 0.01 Ry
was applied to all the compounds.
To estimate $T_c$, we used Allen-Dynes formula~\cite{1957BAR,1965ALL} 
implemented in Quantum Espresso,~\cite{2009PAO} with the effective 
Coulomb interaction $\mu ^*$, being chosen $0.1$ empirically
(See Supplementary Note 3).

\section{References}
\label{sec:ref}


\section{acknowledgments}
\label{sec:acknowledgments}
The authors are grateful to Tomohiro Ichibha for assistance in DFT calculation, 
Takeshi Yajima for useful discussion, and Hiroshi Kageyama for his encouragement.
The authors also acknowledge the support by the Computational 
Materials Science Initiative (CMSI/Japan) for the computational resources, 
Project Nos. hp120086, hp140150, hp150014 at K-computer, and SR16000 
(Center for Computational Materials Science of the 
Institute for Materials Research, Tohoku University/Japan).
R.M. is grateful for financial support from MEXT-KAKENHI
grants 26287063, 25600156, 22104011 and that from the Asahi glass Foundation.
K.H. is grateful for financial support from MEXT-KAKENHI
grants 15K21023 and 15H02672.
The computation in this work has been partially 
performed using the facilities of the Center for Information Science in JAIST.

\section{Author contributions}
\label{sec:author_contribution}

K.N. initiated and performed main calculations under the supervision by R.M.
Data is analysed by K.N. and K.H.
All the authors wrote the paper, section by section, 
finally organized to a manuscript.

\section{Additional information}
Competing financial interests: The authors declare no competing financial interests.

\clearpage
\section{Supplementary Information}
\label{sec:supplementary}
\renewcommand{\refname}{}
\makeatletter
\def\@biblabel#1{[S{#1}]}
\makeatother

\setcounter{figure}{0}
\renewcommand{\thetable}{S-\Roman{table}}
\renewcommand{\thefigure}{S-\arabic{figure}}

\defcitealias{S2013SUE}{S1}
\defcitealias{S2013SUE2}{S2}
\defcitealias{S2012OUM}{S3}
\defcitealias{S2010WAN}{S4}
\defcitealias{S2012YAJ}{S5}
\defcitealias{S2013YAJ1}{S6}
\defcitealias{S2009PAO}{S7}
\defcitealias{S2013SUB}{S8}
\defcitealias{S1957BAR}{S9}
\defcitealias{S1965ALL}{S10}
\defcitealias{S1960ELI}{S11}
\defcitealias{S2014FRA}{S12}
\defcitealias{S2013YAJ2}{S13}


\subsection{Supplementary Note 1}
\label{sec:sup_note_1}

Our optimized geometry parameters for undistorted 
structures are given in Table~\ref{undistorted_vc_relax}, compared with experimental values.
The optimizations were performed under a fixed symmetry, $P$4/$mmm$,
to relax both lattice parameters, $a$ and $c$, and internal coordinations
within a primitive cell.
For $a$ and the interanal coordinations, our results are
in good agreements with experiments, while those for $c$ are
slightly longer than the experimental values.
This trend is also reported in
previous calculations by Suetin {\it et al.}~{[\citetalias{S2013SUE,S2013SUE2}]}.
This is due to our choice of GGA-PBE, which is known to
overestimate lattice parameters in general~{[\citetalias{S2012OUM}]}.

\begin{table}[htbp]
  \caption{
    Optimized lattice constants, $a$ $(=b)$ and $c$, and
    $z$-components of $Pn$ atomic positions    
    of 
    BaTi$_2${\it Pn}$_2$O ($Pn =$ As, Sb, Bi) under $P4/mmm$ symmetry, 
    compared with experiments and other DFT results.
    All units are given in $\AA$.}
\label{undistorted_vc_relax}  
\begin{center}
\begin{tabular}{cccc}
\hline
BaTi$_2$As$_2$O & $a $ & $c$ & As z-pos. \\
\hline
GGA-PBE (previous)~{[\citetalias{S2013SUE2}]} & 4.057 & 7.263 & 0.2427 \\
GGA-PBE (present) & 4.058 & 7.393 & 0.2422 \\
Experiment~{[\citetalias{S2010WAN}]} & 4.046 & 7.272 & 0.2440 \\
\hline
\hline
BaTi$_2$Sb$_2$O & $a$ & $c$ & Sb z-pos. \\
\hline
GGA-PBE (previous)~{[\citetalias{S2013SUE2}]} & 4.116  & 8.107 & 0.2467\\
GGA-PBE (present) & 4.089 & 8.285 & 0.2451 \\
Experiment~{[\citetalias{S2012YAJ}]} & 4.110 & 8.086 & 0.2487 \\ 
\hline
\hline
BaTi$_2$Bi$_2$O & $a$ & $c$ & Bi z-pos. \\
\hline
GGA-PBE (previous)~{[\citetalias{S2013SUE}]} & 4.122 & 8.547 & 0.2523\\
GGA-PBE (present) & 4.118 & 8.630 & 0.2481 \\
Experiment~{[\citetalias{S2013YAJ1}]} & 4.123 & 8.345 & 0.2513 \\
\hline
\end{tabular}
\end{center}
\end{table}

\subsection{Supplementary Note 2}
\label{sec:sup_note_2}

\begin{figure}[htbp]
  \centering
  \includegraphics[scale=0.5]{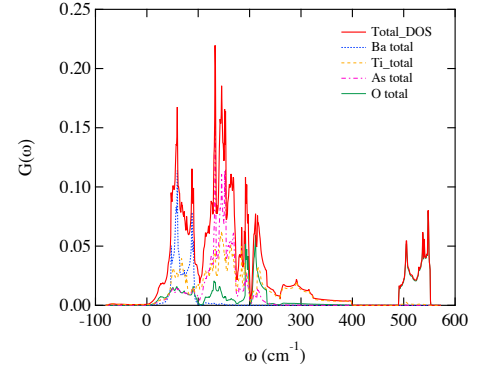}
  \includegraphics[scale=0.5]{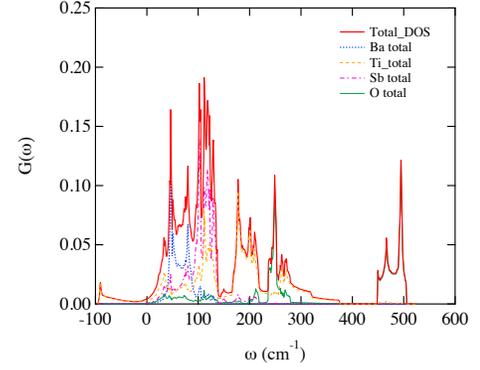}
  \includegraphics[scale=0.5]{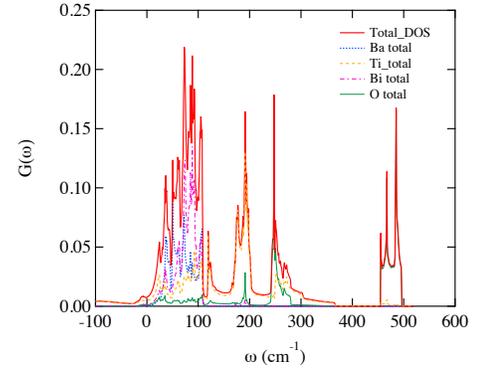}
  \caption{
    Total and partial phonon density of states of BaTi$_2${\it Pn}$_2$O
    under $P$4/$mmm$ symmetry 
    ({\it Pn} = (a) As, (b) Sb, and (c) Bi).}
  \label{btao_phdos1}
\end{figure}

\begin{figure}[htbp]
  \centering
  \includegraphics[scale=0.5]{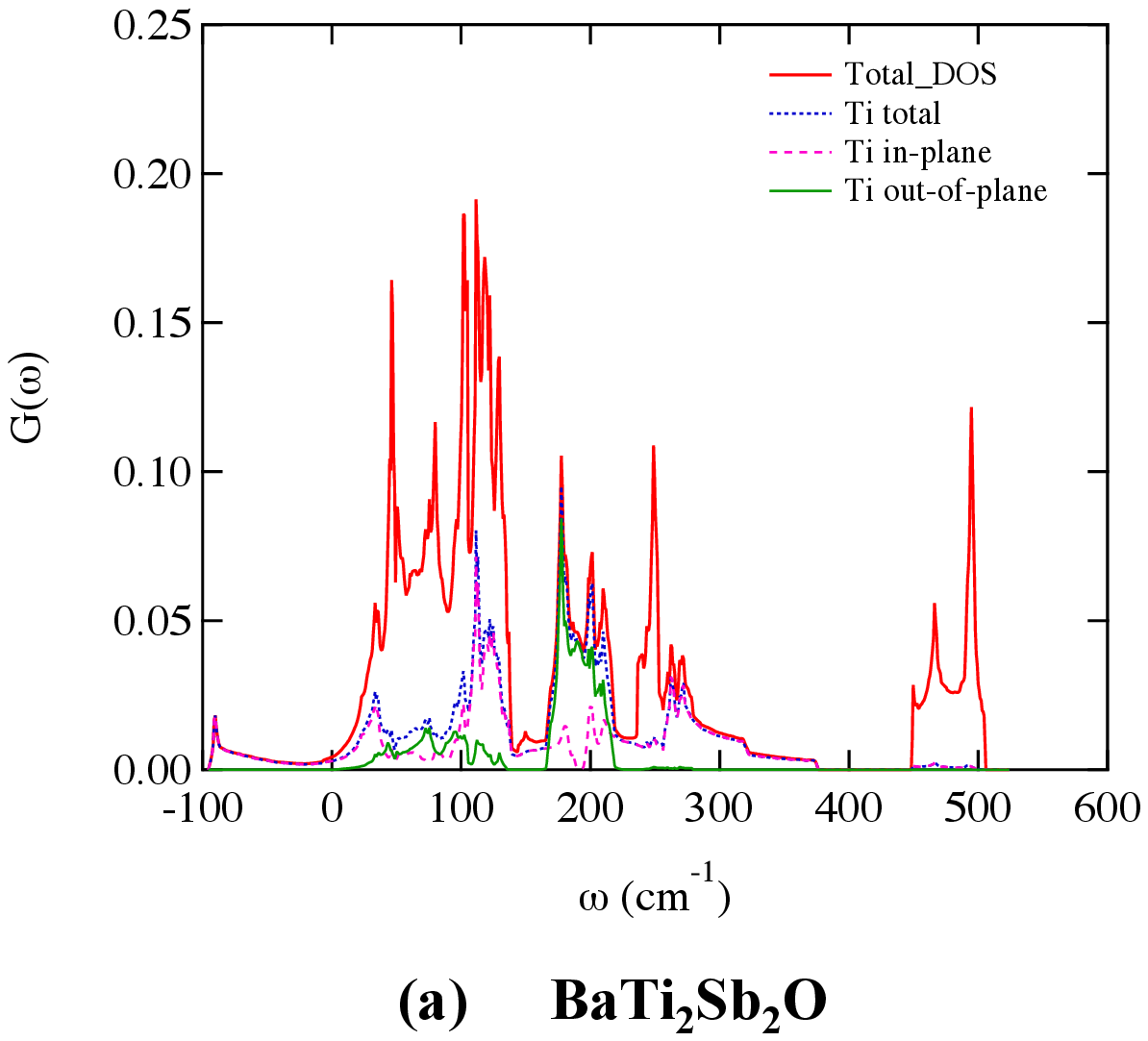}
  \includegraphics[scale=0.5]{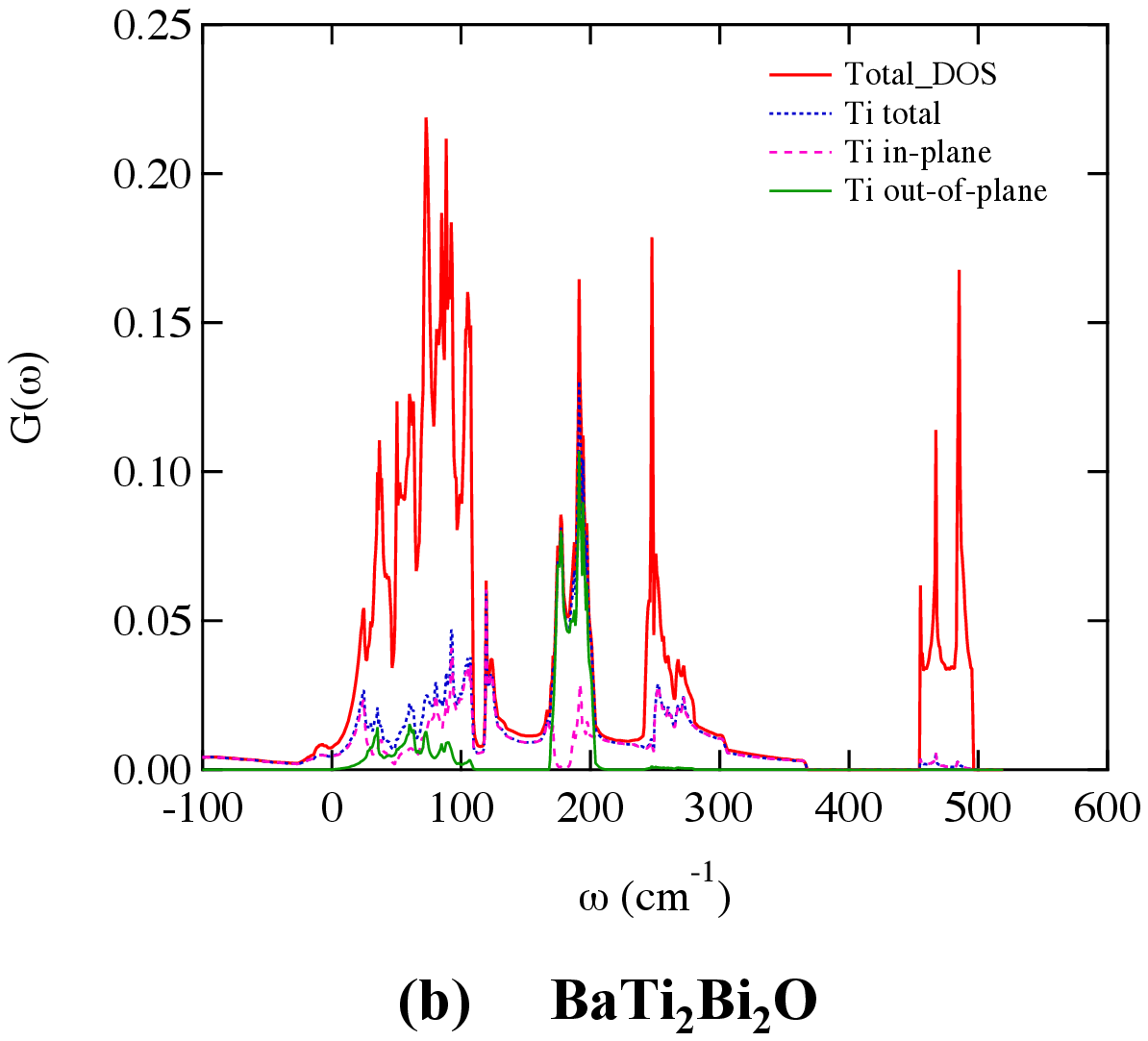}
  \caption{
    Partial density of states for phonons in
    BaTi$_2${\it Pn}$_2$O ({\it Pn} = Sb and Bi),
    divided into the contributions 
    from in-plane and out-of-plane vibrations of Ti atoms.
  }
  \label{btso_phdos2}
\end{figure}

\begin{figure}[htbp]
  \centering
  \includegraphics[scale=0.5]{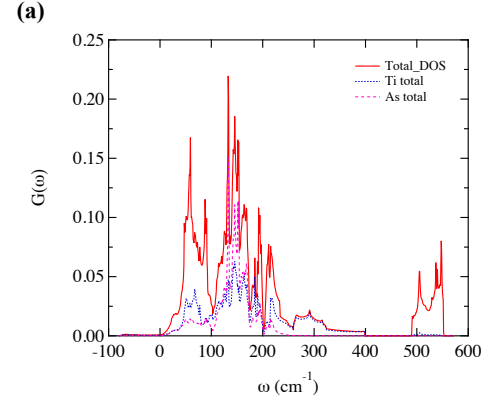}
  \includegraphics[scale=0.5]{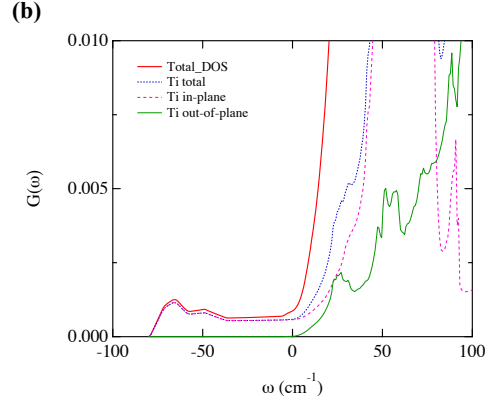}
  \includegraphics[scale=0.5]{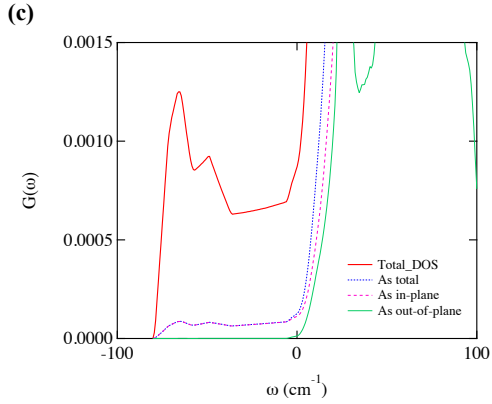}
  \caption{
    Partial density of states for phonons in BaTi$_2$As$_2$O,
    divided into the contributions 
    from in-plane and out-of-plane vibrations of Ti and As atoms.
  }
  \label{btao_phdos3}
\end{figure}

From phonon pDOS (partical DOS), we can identify which
vibration modes lead to the instability toward the superlattice.
We got pDOS with using QHA module which is implemented in Quantum espresso.~{[\citetalias{S2009PAO}]}
For undistorted $Pn$=Sb and Bi, it is found from Figs.~\ref{btao_phdos1} and~\ref{btso_phdos2} that
the negative (imaginary) frequencies mainly come from
Ti 'in-plane' (within $xy$ plane) vibrations.
This is consistent with the previous calculation by Subedi~{[\citetalias{S2013SUB}]} for $Pn$=Sb.
It is found from Fig.~\ref{btao_phdos3} that 
the negative frequencies of undistorted $Pn$=As comes from Ti and As 'in-plane' vibrations. 
Finally, we concluded that only 'in-plane' vibrations contribute to
the negative frequencies for all the compound.


\subsection{Supplementary Note 3}
\label{sec:sup_note_3}
To estimate $T_c$, we used Allen-Dynes formula~{[\citetalias{S1957BAR,S1965ALL}]}
implemented in Quantum Espresso,~{[\citetalias{S2009PAO}]}

\begin{eqnarray}
  {T_c} = \frac{{{\omega _{\ln }}}}{{1.2}}\exp \left[
    {\frac{
        { - 1.04 \left( {1 + \lambda } \right)}
      }{
        { \lambda - {\mu ^*} \left( {1 + 0.62 \lambda } \right)}
      }
    }
    \right] ,
\end{eqnarray}
where
\begin{eqnarray}
    \lambda  = 2\int {d\omega \frac{{{\alpha ^2}
          F\left( \omega  \right)}}{\omega }}
\end{eqnarray}
denotes the frequency-averaged electron-phonon 
coupling constant, and
\begin{eqnarray}
     {\omega _{\ln } }
  = \exp \left[ {\frac{2}{\lambda }\int d\omega \:
      {\alpha ^2}F\left( \omega  \right)\frac{{
          \ln
\omega }}{\omega }} \right]
\end{eqnarray}
denotes logarithm-averaged phonon frequency.
The constant, $\mu ^*$, describes the effective 
Coulomb interaction, being chosen 0.1 empirically.
Eliashberg function {[\citetalias{S1960ELI}]} is given as
\begin{eqnarray}
  {\alpha ^2}F\left( \omega  \right)
  = \frac{1}{{2\pi N\left( {{\varepsilon _F}} \right)}}
  \sum\limits_{q,\nu } {\delta \left( {\omega  - {\omega _{q,\nu }}} \right)} 
  \frac{{{\gamma _{q,\nu }}}}{{\hbar {\omega _{q,\nu }}}} ,
\end{eqnarray}
where $N\left( {{\varepsilon _F}} \right)$, 
$\omega _{q,\nu }$, and $\gamma _{q,\nu }$ denote
the density of state at Fermi-level, 
phonon frequency, and relaxation 
constant for a mode $(q,\nu)$, respectively.

\begin{figure}[ht]
  \centering
  \includegraphics[scale=0.5]{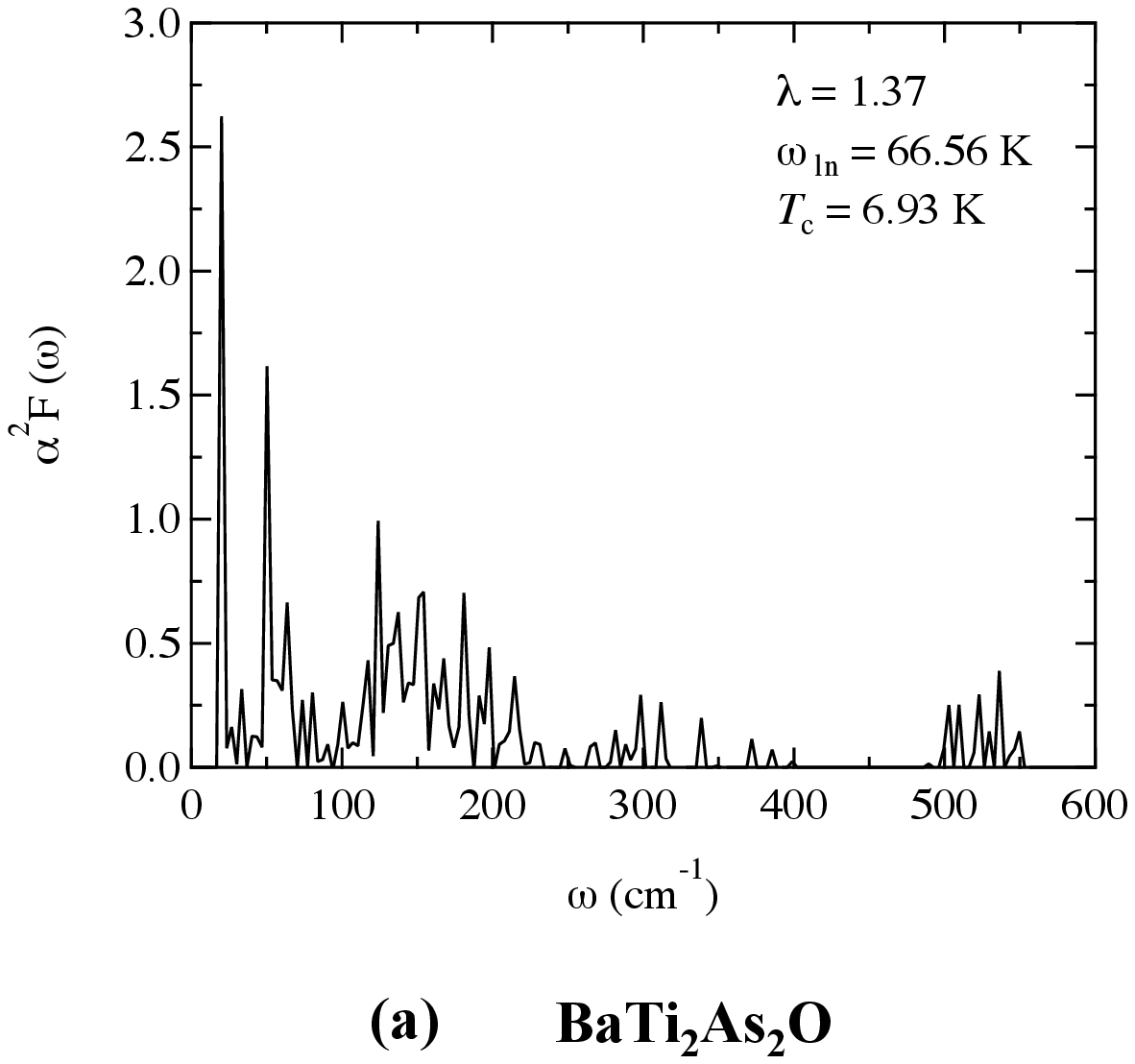}
  \includegraphics[scale=0.5]{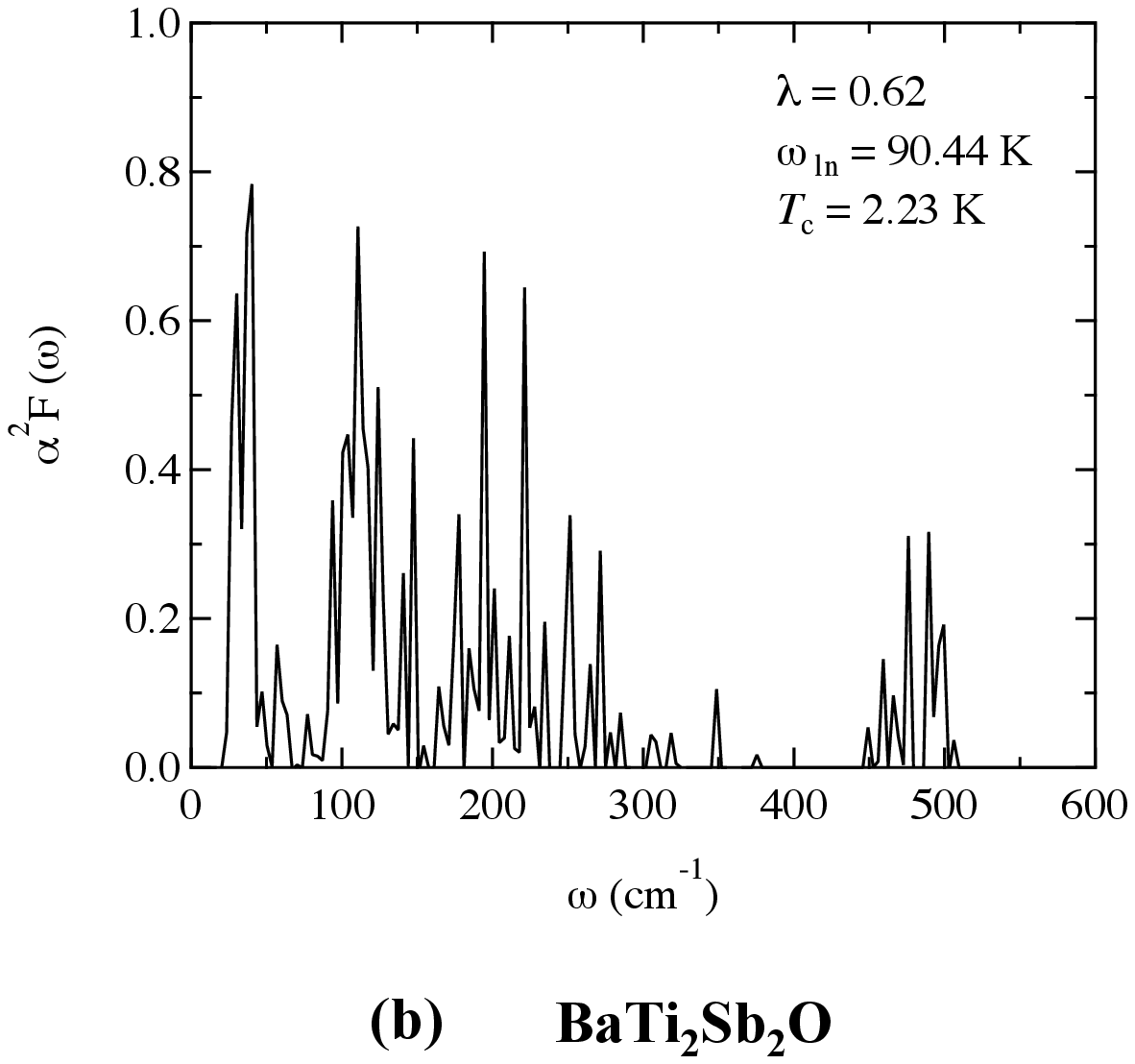}
  \includegraphics[scale=0.5]{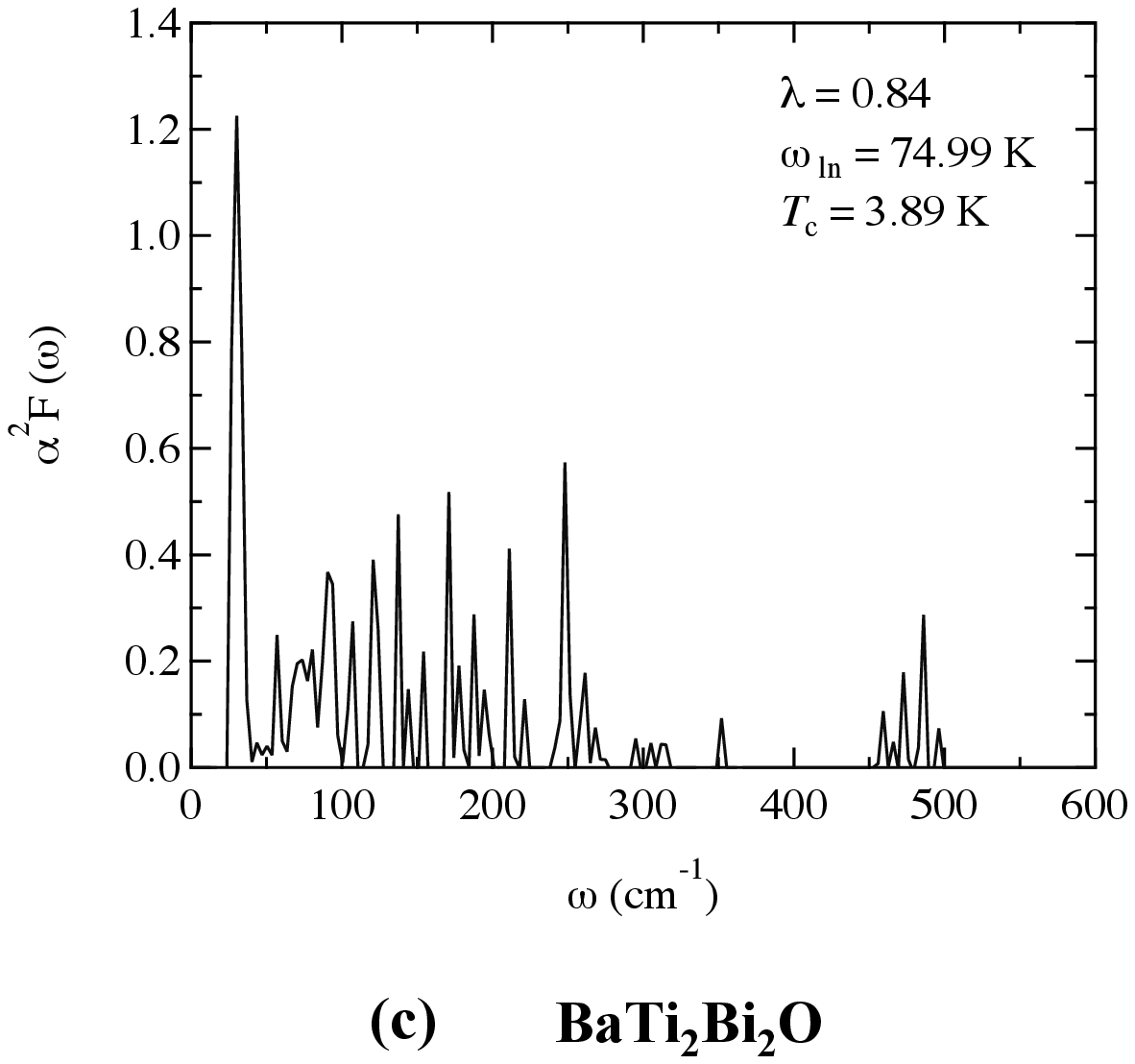}
  \caption{
    Eliashberg spectral function  $\alpha^2 F(\omega)$
    for (a) BaTi$_2$As$_2$O, (b) BaTi$_2$Sb$_2$O and (c) BaTi$_2$Bi$_2$O
    under $P$4/$mmm$ symmetry.
    The imaginary frequencies are not taken into account.
  }
  \label{btso_a2F}
\end{figure}

\begin{figure}[ht]
  \centering
  \includegraphics[scale=0.5]{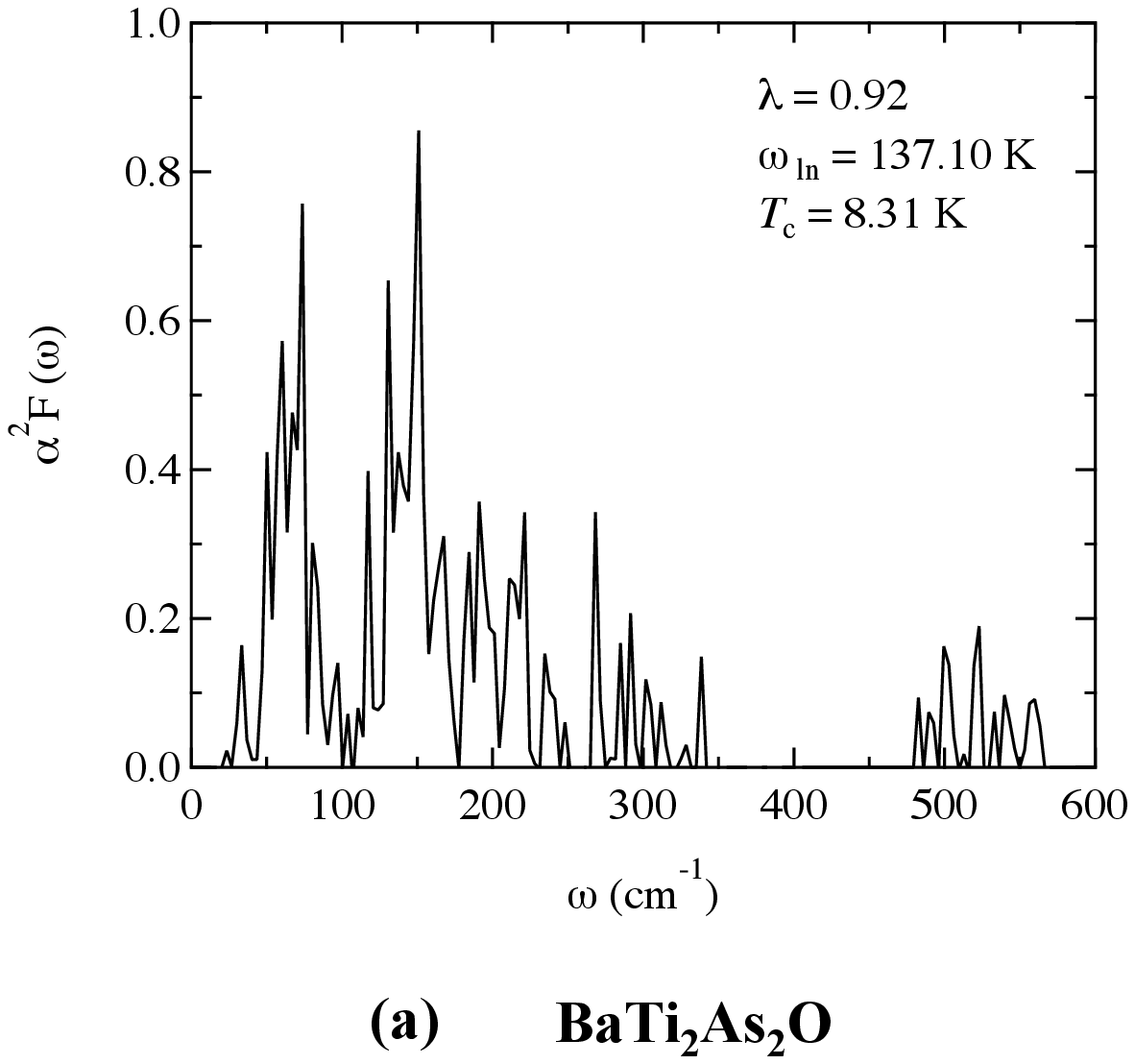}
  \includegraphics[scale=0.5]{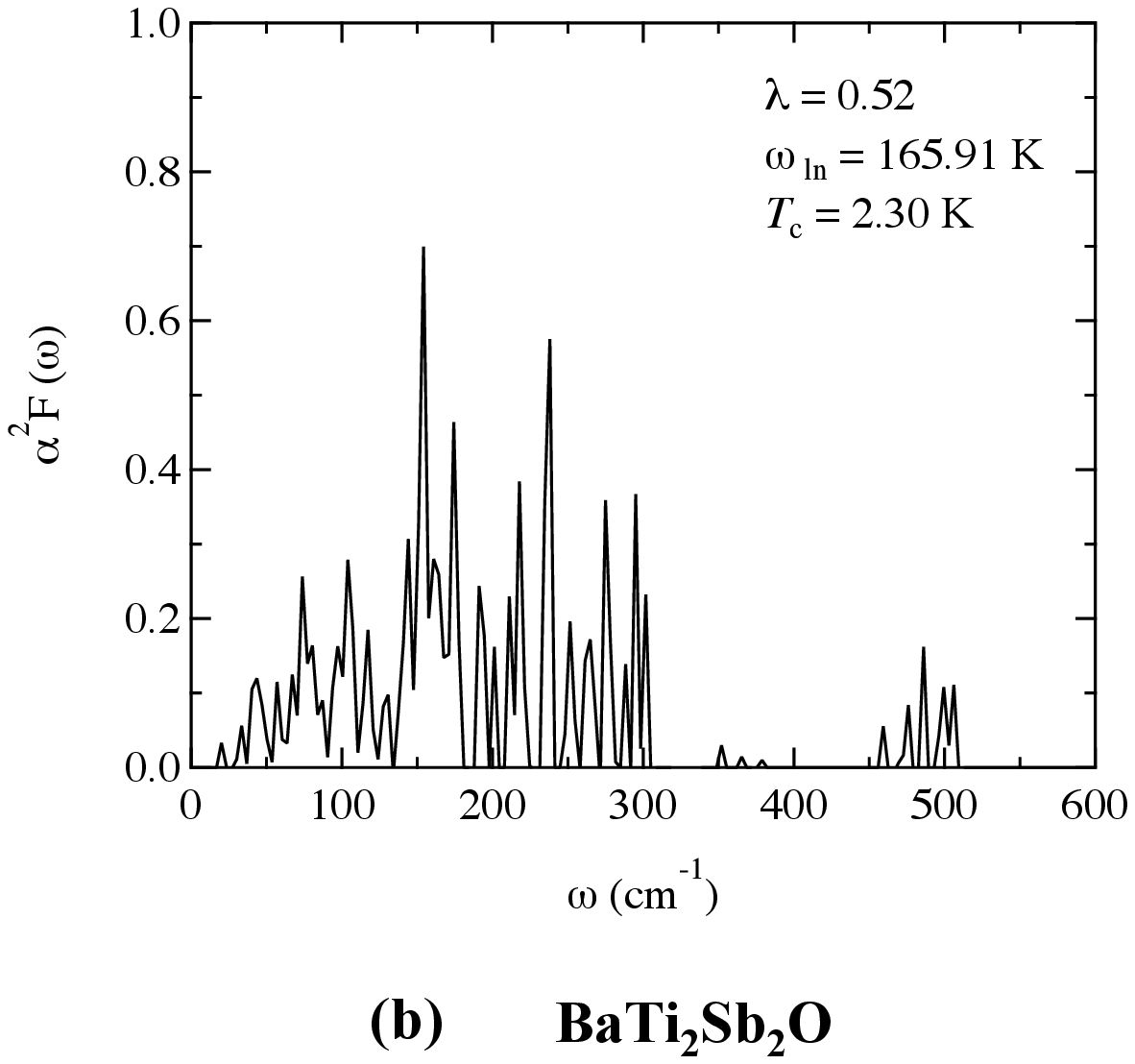}
  \includegraphics[scale=0.5]{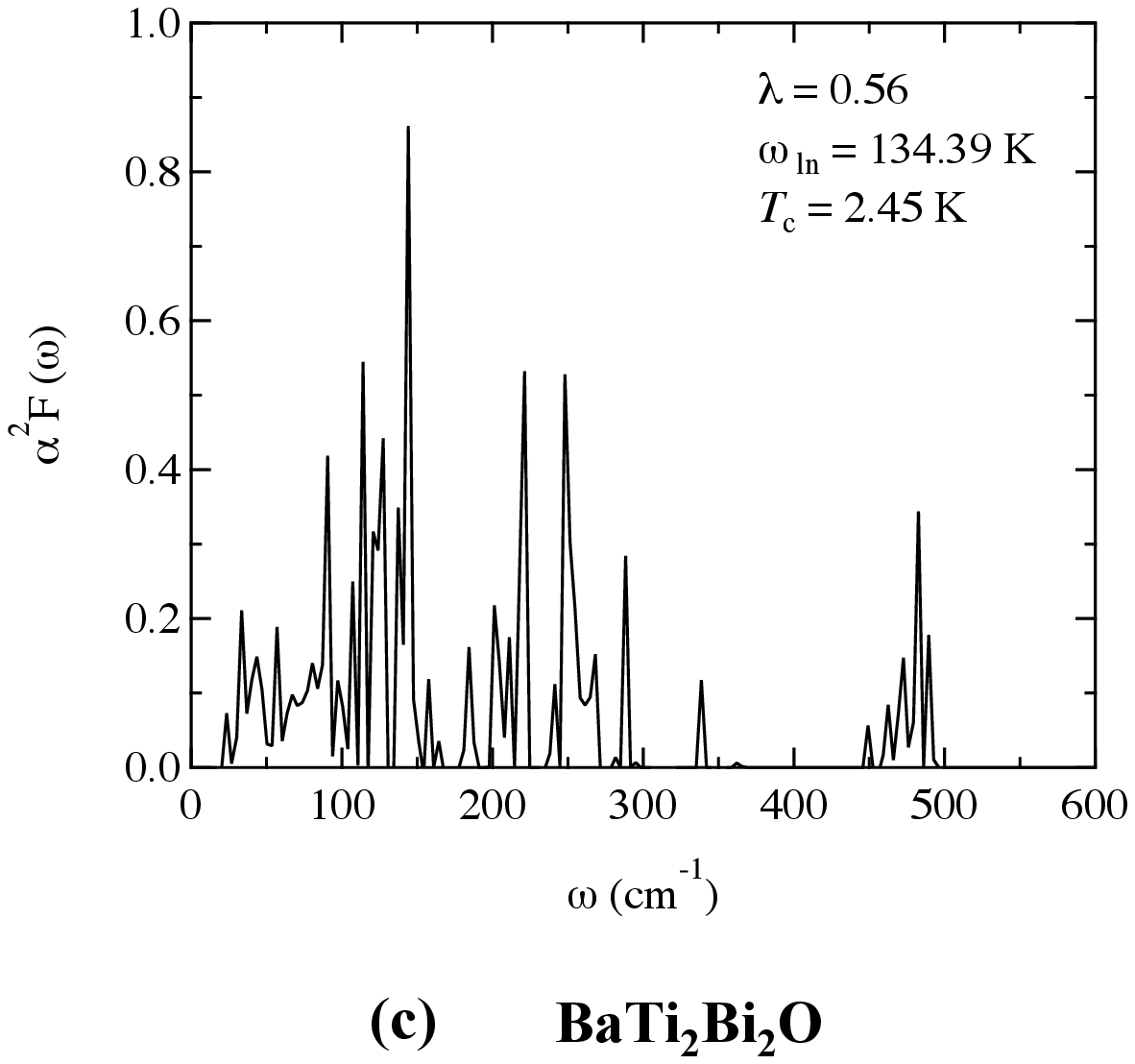}
  \caption{
    Eliashberg spectral function  $\alpha^2 F(\omega)$
    for superlattice structures, (a) BaTi$_2$As$_2$O-$1\times 2 \times 1$, (b) BaTi$_2$Sb$_2$O-$\sqrt{2}\times\sqrt{2}\times 1$ 
	and (c) BaTi$_2$Bi$_2$O-$\sqrt{2}\times\sqrt{2}\times 1$.
    The imaginary frequencies are not taken into account for BaTi$_2$As$_2$O.
  }
  \label{superlattice_a2F}
\end{figure}

\vspace{2mm}
Following the above equation, $T_c$ are estimated for undistorted 
and superlattice structures $Pn$ =As, Sb and Bi cases.
Eliashberg functions are shown in Fig.~\ref{btso_a2F} 
(undistorted  structures) and in Fig.~\ref{superlattice_a2F} (superlattice structures).
The parameters appearing in the formula are 
also tabulated in the table~\ref{table_Tc}.
Our estimated values, $T_c$ = 2.30 (2.45) K for superlattice structures of $Pn$ = Sb (Bi) which show no imaginary frequency, are consistent with experimental values $T_c$ = 1.2 (4.6) K for $Pn$ = Sb (Bi).
This evaluation, however, assumes a simple BCS-type mechanism, 
which might be debatable for BaTi$_2$Bi$_2$O as mentioned in the main article.
As for $Pn$ = As, $T_c$ were estimated only for the unstable structures which
show imaginary frequencies. Therefore, the estimated value $T_c$ = 6.93 K ($1\times 1\times 1$),
8.31 K ($1\times 2\times 1$) are not compatible with the experimental fact that 
BaTi$_2$As$_2$O does not show any superconductivity.~{[\citetalias{S2010WAN, S2013YAJ2}]}


\begin{table*}[htbp]
  \caption{
    $T_c$ obtained by Allen-Dynes formula for BaTi$_2${\it Pn}$_2$O ({\it Pn} = As, Sb, and Bi)}
\label{table_Tc}  
\begin{center}
\begin{tabular}{c|ccc|ccc|c}
 \multicolumn{1}{c|}{}
 & \multicolumn{3}{c|}{Present calculations} & \multicolumn{3}{c|}{Previous calculations} & 
 \multicolumn{1}{c}{Experiments}\\
 \hline
 \hline
 Compounds (structures) & $\lambda$ & $\omega_{ln}$ & $T_c$
 & $\lambda$& $\omega_{ln}$ & $T_c$
 & $T_c$\\
\hline
BaTi$_2$As$_2$O ($1\times 1\times 1$) & 1.37 & 66.56 K & 6.93 K
& - & - & - & -~{[\citetalias{S2010WAN,S2013YAJ2}]} \\
\hline
BaTi$_2$Sb$_2$O ($1\times 1\times 1$) & 0.62 & 90.44 K & 2.23 K 
& 1.28 {[\citetalias{S2013SUB}]} & 93.52 K {[\citetalias{S2013SUB}]} & 9.0 K {[\citetalias{S2013SUB}]} 
& 1.2 K {[\citetalias{S2012YAJ}]}\\
\hline
BaTi$_2$Bi$_2$O ($1\times 1\times 1$) & 0.84 & 74.99 K & 3.89 K 
& - & - & - & 4.6 K {[\citetalias{S2013YAJ1}]}\\
\hline
\hline
BaTi$_2$As$_2$O ($1\times 2\times 1$) & 0.92 & 137.10 K & 8.31 K 
& - & - & - & -~{[\citetalias{S2010WAN,S2013YAJ2}]}\\
\hline
BaTi$_2$Sb$_2$O ($\sqrt{2}\times\sqrt{2}\times 1$) & 0.52 & 165.91 K & 2.30 K 
& 0.55 {[\citetalias{S2013SUB}]} & 110 K {[\citetalias{S2013SUB}]} & 2.7 K {[\citetalias{S2013SUB}]} 
& 1.2 K {[\citetalias{S2012YAJ}]}\\
\hline
BaTi$_2$Bi$_2$O ($\sqrt{2}\times\sqrt{2}\times 1$) & 0.56 & 134.39 K & 2.45 K 
& - & - & - & 4.6 K {[\citetalias{S2013YAJ1}]}\\
\hline
\hline
\end{tabular}
\end{center}
\end{table*}

\subsection{Supplementary Note 4}
\label{sec:sup_note_4}

In general, we can predict superlattice structures by analyzing dynamical matrices.
Once we identify the symmetries, we can perform geometry
optimizations for the superlattice under the identified symmetries and
get relaxed geometry of the superlattice structures.
We found that the lattice instabilities in $Pn$ = Sb and Bi induce structural
transition from $P$4/$mmm$ (No.123) to $P$4/$mbm$ (No.127) by analyzing dynamical matrices. 
The results of geometry optimization under the identified symmetry ($P$4/$mbm$) 
are shown in Tables~\ref{distorted_btso_vc_relax} and~\ref{distorted_btbo_vc_relax}.
On the other hand, we found that the lattice instabilities in $Pn$ = As induce structural
transition from $P$4/$mmm$ (No.123) to $Pbmm$ (No.51), which is different from $Pn$=Sb and Bi.
The results of geometry optimization under the identified symmetry ($Pbmm$)
are summarized in Table~\ref{distorted_btao_vc_relax}.
As mentioned in the main article, the superlattice structure of $Pn$ = As 
($1\times 2\times 1$) still shows imaginary frequencies. We, therefore, further analyzed 
dynamical matrices and found the structural transition from $Pbmm$ (No.51) to $Pbam$ (No.55).
The results of geometry optimization under the identified symmetry ($Pbam$) are summarized in 
Table~\ref{2_2_1_btao_vc_relax}. The results in Tables~\ref{distorted_btao_vc_relax}-\ref{2_2_1_btao_vc_relax}
are compared with experiments and discussed in the main article.


\begin{table}[htbp]
  \caption{
    Internal atomic positions of BaTi$_2$As$_2$O superlattice structure
	($1 \times 2 \times 1$, No.51 $Pbmm$), given in factional coordinates
    for the lattice constants, $a = 4.060$ $\AA$,
    $b = 8.110$ $\AA$, and $c = 7.401$ $\AA$.
    All the atomic positions and lattice constants are simultaneously
    optimized by GGA-PBE.    
    The magnitude of distortions in terms of the orthogonal index
    is evaluated as,
    $\eta = 2 \times (a-1/2b)/(a+1/2b) = 0.115 \%$,
    being fairly coincidence with the experimental value from
    neutron diffraction, {[\citetalias{S2014FRA}]} $\eta = 0.22 \%$.}
  \label{distorted_btao_vc_relax}  
\begin{center}
\begin{tabular}{ccccc}
\hline
Atom & Site & $x$ & $y$ & $z$\\
\hline
Ba & 2$d$ & 0.5000 & 0.5000 & 0.5000\\
Ti & 2$a$ & 0.0000 & 0.0000 & 0.0000\\
Ti & 2$e$ & 0.4851 & 0.2500 & 0.0000\\
As & 4$k$ & 0.0040 & 0.2500 & 0.2422\\
O  & 2$c$ & 0.5000 & 0.0000 & 0.0000\\
\hline
\end{tabular}
\end{center}
\end{table}
\begin{table}[htbp]
  \caption{
    Internal atomic positions of BaTi$_2$Sb$_2$O superlattice structure 
    ($\sqrt{2}\times\sqrt{2}\times 1$, No.127 $P$4/$mbm$), given in factional coordinates
    for the lattice constants, $a = b = 5.791$ $\AA$ and
    $c = 8.349$ $\AA$. 
    All the atomic positions and lattice constants are simultaneously
    optimized by GGA-PBE.    
    Ti atom displaces by 0.14 $\AA$ from its original high-symmetric
    position.}
\label{distorted_btso_vc_relax}  
\begin{center}
\begin{tabular}{ccccc}
\hline
Atom & Site & $x$ & $y$ & $z$\\
\hline
Ba & 2$b$ & 0.5000 & 0.5000 & 0.5000\\
Ti & 4$g$ & 0.7326 & 0.2674 & 0.0000\\
Sb & 4$f$ & 0.0000 & 0.5000 & 0.2460\\
O  & 2$a$ & 0.0000 & 0.0000 & 0.0000\\
\hline
\end{tabular}
\end{center}
\end{table}
\begin{table}[htbp]
  \caption{
    Internal atomic positions of BaTi$_2$Bi$_2$O superlattice structure 
	($\sqrt{2}\times\sqrt{2}\times1$, No.127 $P$4/$mbm$), given in factional coordinates
    for the lattice constants, $a = b = 5.808$ $\AA$ and
    $c = 8.687$ $\AA$. 
    All the atomic positions and lattice constants are simultaneously
    optimized by GGA-PBE.    
    Ti atom displaces by 0.16 $\AA$ from its original high-symmetric
    position.}
\label{distorted_btbo_vc_relax}  
\begin{center}
\begin{tabular}{ccccc}
\hline
Atom & Site & $x$ & $y$ & $z$\\
\hline
Ba & 2$b$ & 0.5000 & 0.5000 & 0.5000\\
Ti & 4$g$ & 0.7309 & 0.2691 & 0.0000\\
Bi & 4$f$ & 0.0000 & 0.5000 & 0.2489\\
O  & 2$a$ & 0.0000 & 0.0000 & 0.0000\\
\hline
\end{tabular}
\end{center}
\end{table}

\begin{table}[htbp]
  \caption{
    Internal atomic positions of BaTi$_2$As$_2$O superlattice structure
    ($2 \times 2 \times 1$, No.55 $Pbam$), given in factional coordinates
    for the lattice constants, $a = 8.122$ $\AA$,
    $b = 8.108$ $\AA$, and $c = 7.401$ $\AA$.
    All the atomic positions and lattice constants are simultaneously
    optimized by GGA-PBE.    
    The magnitude of distortions in terms of the orthorhombicity
    is evaluated as,
    $\eta = 2 \times (a-b)/(a+b) = 0.171 \%$,
    being coincidence with the experimental value from
    neutron diffraction, {[\citetalias{S2014FRA}]} $\eta = 0.22 \%$.}
  \label{2_2_1_btao_vc_relax}  
\begin{center}
\begin{tabular}{ccccc}
\hline
Atom & Site & $x$ & $y$ & $z$\\
\hline
Ba & 2$a$ & 0.5000 & 0.5000 & 0.5000\\
Ba & 2$b$ & 0.0000 & 0.5000 & 0.5000\\
Ti & 4$g$ & 0.2500 & 0.5034 & 0.0000\\
Ti & 4$g$ & 0.0080 & 0.2500 & 0.0000\\
As & 8$i$ & 0.2519 & 0.2509 & 0.2422\\
O  & 2$a$ & 0.0000 & 0.5000 & 0.0000\\
O  & 2$c$ & 0.5000 & 0.0000 & 0.0000\\
\hline
\end{tabular}
\end{center}
\end{table}


\begin{thebibliography}{10}
\expandafter\ifx\csname url\endcsname\relax
  \def\url#1{\texttt{#1}}\fi
\expandafter\ifx\csname urlprefix\endcsname\relax\def\urlprefix{URL }\fi
\providecommand{\bibinfo}[2]{#2}
\providecommand{\eprint}[2][]{\url{#2}}

\bibitem{1997AXE}
\bibinfo{author}{III, E.~A.}, \bibinfo{author}{Ozawa, T.},
  \bibinfo{author}{Kauzlarich, S.~M.} \& \bibinfo{author}{Singh, R.~R.}
\newblock \bibinfo{title}{{Phase Transition and Spin-gap Behavior in a Layered
  Tetragonal Pnictide Oxide}}.
\newblock \emph{\bibinfo{journal}{Journal of Solid State Chemistry}}
  \textbf{\bibinfo{volume}{134}}, \bibinfo{pages}{423--426}
  (\bibinfo{year}{1997}).

\bibitem{2001OZA}
\bibinfo{author}{Ozawa, T.~C.}, \bibinfo{author}{Kauzlarich, S.~M.},
  \bibinfo{author}{Bieringer, M.},  \& \bibinfo{author}{Greedan, J.~E.}
\newblock \bibinfo{title}{{Possible Charge-Density-Wave/Spin-Density-Wave in
  the Layered Pnictide-Oxides:{Na}$_2${Ti}$_2$${Pn}_2${O} (${Pn}$ = {As},
  {Sb})}}.
\newblock \emph{\bibinfo{journal}{Chemistry of Materials}}
  \textbf{\bibinfo{volume}{13}}, \bibinfo{pages}{1804--1810}
  (\bibinfo{year}{2001}).

\bibitem{2009LIU}
\bibinfo{author}{Liu, R.~H.} \emph{et~al.}
\newblock \bibinfo{title}{Physical properties of the layered pnictide oxides
  {Na}$_{2}${Ti}$_{2}{P}_{2}${O} (${P}$={As}, {Sb})}.
\newblock \emph{\bibinfo{journal}{Phys. Rev. B}} \textbf{\bibinfo{volume}{80}},
  \bibinfo{pages}{144516} (\bibinfo{year}{2009}).

\bibitem{2010WAN}
\bibinfo{author}{Wang, X.~F.} \emph{et~al.}
\newblock \bibinfo{title}{Structure and physical properties for a new layered
  pnictide-oxide: {Ba}{Ti}$_2${As}$_2${O}}.
\newblock \emph{\bibinfo{journal}{Journal of Physics: Condensed Matter}}
  \textbf{\bibinfo{volume}{22}}, \bibinfo{pages}{075702}
  (\bibinfo{year}{2010}).

\bibitem{2012YAJ}
\bibinfo{author}{Yajima, T.} \emph{et~al.}
\newblock \bibinfo{title}{Superconductivity in {Ba}{Ti}$_2${Sb}$_2${O} with a
  $d^1$ square lattice}.
\newblock \emph{\bibinfo{journal}{Journal of the Physical Society of Japan}}
  \textbf{\bibinfo{volume}{81}}, \bibinfo{pages}{103706}
  (\bibinfo{year}{2012}).

\bibitem{2012DOA}
\bibinfo{author}{Doan, P.} \emph{et~al.}
\newblock \bibinfo{title}{{Ba}$_{1-x}${Na}$_x${Ti}$_2${Sb}$_2${O} ($0.0 \leq x
  \leq 0.33$) {A Layered Titanium-Based Pnictide Oxide Superconductor}}.
\newblock \emph{\bibinfo{journal}{Journal of the American Chemical Society}}
  \textbf{\bibinfo{volume}{134}}, \bibinfo{pages}{16520--16523}
  (\bibinfo{year}{2012}).

\bibitem{2013YAJ1}
\bibinfo{author}{Yajima, T.} \emph{et~al.}
\newblock \bibinfo{title}{{Synthesis and Physical Properties of the New
  Oxybismuthides {Ba}{Ti}$_2${Bi}$_2${O} and ({SrF})$_2${Ti}$_2${Bi}$_2${O}
  with a $d^1$ Square Net}}.
\newblock \emph{\bibinfo{journal}{Journal of the Physical Society of Japan}}
  \textbf{\bibinfo{volume}{82}}, \bibinfo{pages}{013703}
  (\bibinfo{year}{2013}).

\bibitem{2013YAJ2}
\bibinfo{author}{Yajima, T.} \emph{et~al.}
\newblock \bibinfo{title}{{Two Superconducting Phases in the Isovalent Solid
  Solutions {Ba}{Ti}$_2$${Pn}_2${O} (${Pn}$ = {As}, {Sb}, and {Bi})}}.
\newblock \emph{\bibinfo{journal}{Journal of the Physical Society of Japan}}
  \textbf{\bibinfo{volume}{82}}, \bibinfo{pages}{033705}
  (\bibinfo{year}{2013}).

\bibitem{2013ZHA}
\bibinfo{author}{Zhai, H.-F.} \emph{et~al.}
\newblock \bibinfo{title}{Superconductivity, charge- or spin-density wave, and
  metal-nonmetal transition in
  {Ba}{Ti}$_{2}$({Sb}$_{1-x}${Bi}$_{x}$)$_{2}${O}}.
\newblock \emph{\bibinfo{journal}{Phys. Rev. B}} \textbf{\bibinfo{volume}{87}},
  \bibinfo{pages}{100502} (\bibinfo{year}{2013}).

\bibitem{2013NAK}
\bibinfo{author}{Nakano, K.} \emph{et~al.}
\newblock \bibinfo{title}{{${T}_c$ Enhancement by Aliovalent Anionic
  Substitution in Superconducting {Ba}{Ti}$_2$({Sb}$_{1-x}${Sn}$_x$)$_2${O}}}.
\newblock \emph{\bibinfo{journal}{Journal of the Physical Society of Japan}}
  \textbf{\bibinfo{volume}{82}}, \bibinfo{pages}{074707}
  (\bibinfo{year}{2013}).

\bibitem{2014PAC}
\bibinfo{author}{Pachmayr, U.} \& \bibinfo{author}{Johrendt, D.}
\newblock \bibinfo{title}{Superconductivity in
  {Ba}$_{1-x}${K}$_{x}${Ti}$_{2}${Sb}$_{2}${O} ($0 \leqq x \leqq 1$) controlled
  by the layer charge}.
\newblock \emph{\bibinfo{journal}{Solid State Sciences}}
  \textbf{\bibinfo{volume}{28}}, \bibinfo{pages}{31--34}
  (\bibinfo{year}{2014}).

\bibitem{2014ROH}
\bibinfo{author}{von Rohr, F.}, \bibinfo{author}{Nesper, R.} \&
  \bibinfo{author}{Schilling, A.}
\newblock \bibinfo{title}{Superconductivity in rubidium-substituted
  {Ba}$_{1-x}${Rb}$_{x}${Ti}$_{2}${Sb}$_{2}${O}}.
\newblock \emph{\bibinfo{journal}{Phys. Rev. B}} \textbf{\bibinfo{volume}{89}},
  \bibinfo{pages}{094505} (\bibinfo{year}{2014}).

\bibitem{1957BAR}
\bibinfo{author}{Bardeen, J.}, \bibinfo{author}{Cooper, L.~N.} \&
  \bibinfo{author}{Schrieffer, J.~R.}
\newblock \bibinfo{title}{Theory of superconductivity}.
\newblock \emph{\bibinfo{journal}{Phys. Rev.}} \textbf{\bibinfo{volume}{108}},
  \bibinfo{pages}{1175} (\bibinfo{year}{1957}).

\bibitem{1965ALL}
\bibinfo{author}{Allen, P.~B.} \& \bibinfo{author}{Dynes, R.~C.}
\newblock \bibinfo{title}{Transition temperature of strong-coupled
  superconductors reanalyzed}.
\newblock \emph{\bibinfo{journal}{Phys. Rev. B}} \textbf{\bibinfo{volume}{12}},
  \bibinfo{pages}{905--922} (\bibinfo{year}{1965}).

\bibitem{2013SUB}
\bibinfo{author}{Subedi, A.}
\newblock \bibinfo{title}{Electron-phonon superconductivity and charge density
  wave instability in the layered titanium-based pnictide
  {Ba}{Ti}$_{2}${Sb}$_{2}${O}}.
\newblock \emph{\bibinfo{journal}{Phys. Rev. B}} \textbf{\bibinfo{volume}{87}},
  \bibinfo{pages}{054506} (\bibinfo{year}{2013}).

\bibitem{2013KIT}
\bibinfo{author}{Kitagawa, S.}, \bibinfo{author}{Ishida, K.},
  \bibinfo{author}{Nakano, K.}, \bibinfo{author}{Yajima, T.} \&
  \bibinfo{author}{Kageyama, H.}
\newblock \bibinfo{title}{$s$-wave superconductivity in superconducting
  {Ba}{Ti}$_{2}${Sb}$_{2}${O} revealed by $^{121/123}${Sb}-{NMR}/nuclear
  quadrupole resonance measurements}.
\newblock \emph{\bibinfo{journal}{Phys. Rev. B}} \textbf{\bibinfo{volume}{87}},
  \bibinfo{pages}{060510} (\bibinfo{year}{2013}).

\bibitem{2013NOZ}
\bibinfo{author}{Nozaki, Y.} \emph{et~al.}
\newblock \bibinfo{title}{Muon spin relaxation and electron/neutron diffraction
  studies of {Ba}{Ti}$_{2}$({As}$_{1-x}${Sb}$_{x}$)$_{2}${O}: {Absence of
  static magnetism and superlattice reflections}}.
\newblock \emph{\bibinfo{journal}{Phys. Rev. B}} \textbf{\bibinfo{volume}{88}},
  \bibinfo{pages}{214506} (\bibinfo{year}{2013}).

\bibitem{2013ROH}
\bibinfo{author}{von Rohr, F.}, \bibinfo{author}{Schilling, A.},
  \bibinfo{author}{Nesper, R.}, \bibinfo{author}{Baines, C.} \&
  \bibinfo{author}{Bendele, M.}
\newblock \bibinfo{title}{Conventional superconductivity and
  charge-density-wave ordering in
  {Ba}$_{1-x}${Na}$_{x}${Ti}$_{2}${Sb}$_{2}${O}}.
\newblock \emph{\bibinfo{journal}{Phys. Rev. B}} \textbf{\bibinfo{volume}{88}},
  \bibinfo{pages}{140501} (\bibinfo{year}{2013}).

\bibitem{1986BED}
\bibinfo{author}{Bednorz, J.} \& \bibinfo{author}{M{\"u}ller, K.}
\newblock \bibinfo{title}{Possible high-${T}_c$ superconductivity in the
  {Ba-La-Cu-O} system}.
\newblock \emph{\bibinfo{journal}{Zeitschrift f{\"u}r Physik B Condensed
  Matter}} \textbf{\bibinfo{volume}{64}}, \bibinfo{pages}{189--193}
  (\bibinfo{year}{1986}).

\bibitem{2008KAM}
\bibinfo{author}{Kamihara, Y.}, \bibinfo{author}{Watanabe, T.},
  \bibinfo{author}{Hirano, M.} \& \bibinfo{author}{Hosono, H.}
\newblock \bibinfo{title}{{Iron-Based Layered Superconductor
  {La}[{O}$_{1-x}${F}$_{x}$]{Fe}{As} ($x$ = 0.05 - 0.12) with ${T}_c$ = 26 K}}.
\newblock \emph{\bibinfo{journal}{Journal of the American Chemical Society}}
  \textbf{\bibinfo{volume}{130}}, \bibinfo{pages}{3296--3297}
  (\bibinfo{year}{2008}).

\bibitem{2012SIN}
\bibinfo{author}{Singh, D.~J.}
\newblock \bibinfo{title}{Electronic structure, disconnected {Fermi} surfaces
  and antiferromagnetism in the layered pnictide superconductor
  {Na}$_x${Ba}$_{1-x}${Ti}$_2${Sb}$_2${O}}.
\newblock \emph{\bibinfo{journal}{New Journal of Physics}}
  \textbf{\bibinfo{volume}{14}}, \bibinfo{pages}{123003}
  (\bibinfo{year}{2012}).

\bibitem{2013WAN1}
\bibinfo{author}{Wang, G.}, \bibinfo{author}{Zhang, H.},
  \bibinfo{author}{Zhang, L.} \& \bibinfo{author}{Liu, C.}
\newblock \bibinfo{title}{The electronic structure and magnetism of
  {Ba}{Ti}$_2${Sb}$_2${O}}.
\newblock \emph{\bibinfo{journal}{Journal of Applied Physics}}
  \textbf{\bibinfo{volume}{113}} (\bibinfo{year}{2013}).

\bibitem{2014YU}
\bibinfo{author}{Yu, X.-L.} \emph{et~al.}
\newblock \bibinfo{title}{A site-selective antiferromagnetic ground state in
  layered pnictide-oxide {Ba}{Ti}$_2${As}$_2${O}}.
\newblock \emph{\bibinfo{journal}{Journal of Applied Physics}}
  \textbf{\bibinfo{volume}{115}}, \bibinfo{pages}{17A924}
  (\bibinfo{year}{2014}).

\bibitem{2013SUE}
\bibinfo{author}{Suetin, D.} \& \bibinfo{author}{Ivanovskii, A.}
\newblock \bibinfo{title}{Electronic properties and fermi surface for new
  {Fe}-free layered pnictide-oxide superconductor {Ba}{Ti}$_2${Bi}$_2${O} from
  first principles}.
\newblock \emph{\bibinfo{journal}{JETP Letters}} \textbf{\bibinfo{volume}{97}},
  \bibinfo{pages}{220--225} (\bibinfo{year}{2013}).

\bibitem{2014Xu}
\bibinfo{author}{Xu, H.~C.} \emph{et~al.}
\newblock \bibinfo{title}{Electronic structure of the
  {Ba}{Ti}$_{2}${As}$_{2}${O} parent compound of the titanium-based oxypnictide
  superconductor}.
\newblock \emph{\bibinfo{journal}{Phys. Rev. B}} \textbf{\bibinfo{volume}{89}},
  \bibinfo{pages}{155108} (\bibinfo{year}{2014}).

\bibitem{2016SON}
\bibinfo{author}{Song, Q.} \emph{et~al.}
\newblock \bibinfo{title}{Electronic structure of the titanium-based
  oxypnictide superconductor {Ba}$_{0.95}${Na}$_{0.05}${Ti}$_{2}${Sb}$_{2}${O}
  and direct observation of its charge density wave order}.
\newblock \emph{\bibinfo{journal}{Phys. Rev. B}} \textbf{\bibinfo{volume}{93}},
  \bibinfo{pages}{024508} (\bibinfo{year}{2016}).

\bibitem{2014FRA}
\bibinfo{author}{Frandsen, B.~A.} \emph{et~al.}
\newblock \bibinfo{title}{Intra-unit-cell nematic charge order in the
  titanium-oxypnictide family of superconductors}.
\newblock \emph{\bibinfo{journal}{Nat. Commun.}}
  \textbf{\bibinfo{volume}{5}}, \bibinfo{pages}{5761} (\bibinfo{year}{2014}).

\bibitem{2010LAW}
\bibinfo{author}{Lawler, M.~J.} \emph{et~al.}
\newblock \bibinfo{title}{Intra-unit-cell electronic nematicity of the
  high-${T}_c$ copper-oxide pseudogap states}.
\newblock \emph{\bibinfo{journal}{Nature}} \textbf{\bibinfo{volume}{466}},
  \bibinfo{pages}{347--351} (\bibinfo{year}{2010}).

\bibitem{2014FUJ}
\bibinfo{author}{Fujita, K.} \emph{et~al.}
\newblock \bibinfo{title}{Direct phase-sensitive identification of a $d$-form
  factor density wave in underdoped cuprates}.
\newblock \emph{\bibinfo{journal}{Proceedings of the National Academy of
  Sciences}} \textbf{\bibinfo{volume}{111}}, \bibinfo{pages}{E3026}
  (\bibinfo{year}{2014}).

\bibitem{1991KOI1}
\bibinfo{author}{Koike, Y.}, \bibinfo{author}{Watanabe, N.},
  \bibinfo{author}{Noji, T.} \& \bibinfo{author}{Saito, Y.}
\newblock \bibinfo{title}{Effects of the {Cu}-site substitution on the
  anomalous $x$ dependence of ${T}_c$ in {La}$_{2-x}${Ba}$_x${Cu}{O}$_4$}.
\newblock \emph{\bibinfo{journal}{Solid State Communications}}
  \textbf{\bibinfo{volume}{78}}, \bibinfo{pages}{511--514}
  (\bibinfo{year}{1991}).

\bibitem{1991KOI2}
\bibinfo{author}{Koike, Y.}, \bibinfo{author}{Kawaguchi, T.},
  \bibinfo{author}{Watanabe, N.}, \bibinfo{author}{Noji, T.} \&
  \bibinfo{author}{Saito, Y.}
\newblock \bibinfo{title}{Superconductivity and low-temperature structural
  phase transition in {La}$_{1.98-x}${Ce}$_{0.02}${Ba}$_x${Cu}{O}$_4$}.
\newblock \emph{\bibinfo{journal}{Solid State Communications}}
  \textbf{\bibinfo{volume}{79}}, \bibinfo{pages}{155--158}
  (\bibinfo{year}{1991}).

\bibitem{1991TAM}
\bibinfo{author}{Tamegai, T.} \& \bibinfo{author}{Iye, Y.}
\newblock \bibinfo{title}{Universal transport anomaly in
  {YBa}$_{2}${Cu}$_{3}${O}$_{7}$-type systems with reduced carrier density}.
\newblock \emph{\bibinfo{journal}{Phys. Rev. B}} \textbf{\bibinfo{volume}{44}},
  \bibinfo{pages}{10167} (\bibinfo{year}{1991}).

\bibitem{1992KOI}
\bibinfo{author}{Koike, Y.} \emph{et~al.}
\newblock \bibinfo{title}{Anomalous $x$ dependence of ${T}_c$ and possibility
  of low-temperature structural phase transition in
  {La}$_{2-x}${Sr}$_x${Cu}$_{0.99}$${M}_{0.01}${O}$_4$ (${M}$ = {Ni}, {Zn},
  {Ga})}.
\newblock \emph{\bibinfo{journal}{Solid State Communications}}
  \textbf{\bibinfo{volume}{82}}, \bibinfo{pages}{889--893}
  (\bibinfo{year}{1992}).

\bibitem{2004AND}
\bibinfo{author}{Ando, Y.}, \bibinfo{author}{Komiya, S.},
  \bibinfo{author}{Segawa, K.}, \bibinfo{author}{Ono, S.} \&
  \bibinfo{author}{Kurita, Y.}
\newblock \bibinfo{title}{{Electronic Phase Diagram of High-${T}_{c}$ Cuprate
  Superconductors from a Mapping of the In-Plane Resistivity Curvature}}.
\newblock \emph{\bibinfo{journal}{Phys. Rev. Lett.}}
  \textbf{\bibinfo{volume}{93}}, \bibinfo{pages}{267001}
  (\bibinfo{year}{2004}).

\bibitem{2012LIM}
\bibinfo{author}{Iimura, S.} \emph{et~al.}
\newblock \bibinfo{title}{Two-dome structure in electron-doped iron arsenide
  superconductors}.
\newblock \emph{\bibinfo{journal}{Nat. Commun.}}
  \textbf{\bibinfo{volume}{3}}, \bibinfo{pages}{943} (\bibinfo{year}{2012}).

\bibitem{2014MAT}
\bibinfo{author}{Matsuishi, S.}, \bibinfo{author}{Maruyama, T.},
  \bibinfo{author}{Iimura, S.} \& \bibinfo{author}{Hosono, H.}
\newblock \bibinfo{title}{Controlling factors of ${T}_c$ dome structure in
  1111-type iron arsenide superconductors}.
\newblock \emph{\bibinfo{journal}{Phys. Rev. B}} \textbf{\bibinfo{volume}{89}},
  \bibinfo{pages}{094510} (\bibinfo{year}{2014}).

\bibitem{2013FUJ}
\bibinfo{author}{Fujiwara, N.} \emph{et~al.}
\newblock \bibinfo{title}{{Detection of Antiferromagnetic Ordering in Heavily
  Doped {LaFeAsO}$_{1-x}${H}$_{x}$ Pnictide Superconductors Using
  Nuclear-Magnetic-Resonance Techniques}}.
\newblock \emph{\bibinfo{journal}{Phys. Rev. Lett.}}
  \textbf{\bibinfo{volume}{111}}, \bibinfo{pages}{097002}
  (\bibinfo{year}{2013}).

\bibitem{2013YIL}
\bibinfo{author}{Yildirim, T.}
\newblock \bibinfo{title}{Ferroelectric soft phonons, charge density wave
  instability, and strong electron-phonon coupling in {Bi}{S}$_{2}$ layered
  superconductors: A first-principles study}.
\newblock \emph{\bibinfo{journal}{Phys. Rev. B}} \textbf{\bibinfo{volume}{87}},
  \bibinfo{pages}{020506} (\bibinfo{year}{2013}).

\bibitem{2006JOH}
\bibinfo{author}{Johannes, M.~D.}, \bibinfo{author}{Mazin, I.~I.} \&
  \bibinfo{author}{Howells, C.~A.}
\newblock \bibinfo{title}{Fermi-surface nesting and the origin of the
  charge-density wave in {Nb}{Se}$_{2}$}.
\newblock \emph{\bibinfo{journal}{Phys. Rev. B}} \textbf{\bibinfo{volume}{73}},
  \bibinfo{pages}{205102} (\bibinfo{year}{2006}).

\bibitem{2008JOH}
\bibinfo{author}{Johannes, M.~D.} \& \bibinfo{author}{Mazin, I.~I.}
\newblock \bibinfo{title}{Fermi surface nesting and the origin of charge
  density waves in metals}.
\newblock \emph{\bibinfo{journal}{Phys. Rev. B}} \textbf{\bibinfo{volume}{77}},
  \bibinfo{pages}{165135} (\bibinfo{year}{2008}).

\bibitem{2009CAL}
\bibinfo{author}{Calandra, M.}, \bibinfo{author}{Mazin, I.~I.} \&
  \bibinfo{author}{Mauri, F.}
\newblock \bibinfo{title}{Effect of dimensionality on the charge-density wave
  in few-layer {2H-NbSe}$_{2}$}.
\newblock \emph{\bibinfo{journal}{Phys. Rev. B}} \textbf{\bibinfo{volume}{80}},
  \bibinfo{pages}{241108} (\bibinfo{year}{2009}).

\bibitem{2011CAL}
\bibinfo{author}{Calandra, M.} \& \bibinfo{author}{Mauri, F.}
\newblock \bibinfo{title}{{Charge-Density Wave and Superconducting Dome in
  {TiSe}$_{2}$ from Electron-Phonon Interaction}}.
\newblock \emph{\bibinfo{journal}{Phys. Rev. Lett.}}
  \textbf{\bibinfo{volume}{106}}, \bibinfo{pages}{196406}
  (\bibinfo{year}{2011}).
  
\bibitem{2015ZHU}
\bibinfo{author}{Zhu, X.}, \bibinfo{author}{Cao, Y.}, \bibinfo{author}{Zhang, J.}, 
\bibinfo{author}{Plummer, E. W.}, \& \bibinfo{author}{Guo, J.}
\newblock \bibinfo{title}{Classification of charge density waves based on their nature.}
\newblock \emph{\bibinfo{journal}{Proc. Natl. Acad. Sci. USA}}
  \textbf{\bibinfo{volume}{112}}, \bibinfo{pages}{2367--2371} (\bibinfo{year}{2015}).

\bibitem{2013WAN2}
\bibinfo{author}{Wan, X.}, \bibinfo{author}{Ding, H.-C.},
  \bibinfo{author}{Savrasov, S.~Y.} \& \bibinfo{author}{Duan, C.-G.}
\newblock \bibinfo{title}{Electron-phonon superconductivity near
  charge-density-wave instability in {La}{O}$_{0.5}${F}$_{0.5}${BiS}$_{2}$:
  {Density-functional calculations}}.
\newblock \emph{\bibinfo{journal}{Phys. Rev. B}} \textbf{\bibinfo{volume}{87}},
  \bibinfo{pages}{115124} (\bibinfo{year}{2013}).

\bibitem{2012MIZ}
\bibinfo{author}{Mizuguchi, Y.} \emph{et~al.}
\newblock \bibinfo{title}{Superconductivity in novel {BiS}$_2$-based layered
  superconductor {La}{O}$_{1-x}${F}$_x${Bi}{S}$_2$}.
\newblock \emph{\bibinfo{journal}{Journal of the Physical Society of Japan}}
  \textbf{\bibinfo{volume}{81}}, \bibinfo{pages}{114725}
  (\bibinfo{year}{2012}).
  
\bibitem{2014ARO}
\bibinfo{author}{Aroyo, M. I.} \emph{et~al.}
\newblock \bibinfo{title}{Brillouin-zone databases on the {\it Bilbao Crystallographic Server}}.
\newblock \emph{\bibinfo{journal}{Acta Cryst.}}
  \textbf{\bibinfo{volume}{A70}}, \bibinfo{pages}{126--137}
  (\bibinfo{year}{2014}).

\bibitem{2013LEE}
\bibinfo{author}{Lee, J.} \emph{et~al.}
\newblock \bibinfo{title}{Crystal structure, lattice vibrations, and
  superconductivity of {La}{O}$_{1-x}${F}$_{x}${Bi}{S}$_{2}$}.
\newblock \emph{\bibinfo{journal}{Phys. Rev. B}} \textbf{\bibinfo{volume}{87}},
  \bibinfo{pages}{205134} (\bibinfo{year}{2013}).

\bibitem{2015ATH}
\bibinfo{author}{Athauda, A.} \emph{et~al.}
\newblock \bibinfo{title}{In-plane charge fluctuations in bismuth-sulfide
  superconductors}.
\newblock \emph{\bibinfo{journal}{Phys. Rev. B}} \textbf{\bibinfo{volume}{91}},
  \bibinfo{pages}{144112} (\bibinfo{year}{2015}).

\bibitem{1996PER}
\bibinfo{author}{Perdew, J.~P.}, \bibinfo{author}{Burke, K.} \&
  \bibinfo{author}{Ernzerhof, M.}
\newblock \bibinfo{title}{{Generalized Gradient Approximation Made Simple}}.
\newblock \emph{\bibinfo{journal}{Phys. Rev. Lett.}}
  \textbf{\bibinfo{volume}{77}}, \bibinfo{pages}{3865} (\bibinfo{year}{1996}).

\bibitem{2009PAO}
\bibinfo{author}{Giannozzi, P.} \emph{et~al.}
\newblock \bibinfo{title}{{QUANTUM ESPRESSO}: a modular and open-source
  software project for quantum simulations of materials}.
\newblock \emph{\bibinfo{journal}{Journal of Physics: Condensed Matter}}
  \textbf{\bibinfo{volume}{21}}, \bibinfo{pages}{395502}
  (\bibinfo{year}{2009}).

\bibitem{1994BLO}
\bibinfo{author}{Bl\"ochl, P.~E.}
\newblock \bibinfo{title}{Projector augmented-wave method}.
\newblock \emph{\bibinfo{journal}{Phys. Rev. B}} \textbf{\bibinfo{volume}{50}},
  \bibinfo{pages}{17953} (\bibinfo{year}{1994}).

\bibitem{2014KUC}
\bibinfo{author}{Jollet, F.}, \bibinfo{author}{Torrent, M.} \&
  \bibinfo{author}{Holzwarth, N.}
\newblock \bibinfo{title}{Generation of projector augmented-wave atomic data:
  {A} 71 element validated table in the {XML} format}.
\newblock \emph{\bibinfo{journal}{Computer Physics Communications}}
  \textbf{\bibinfo{volume}{185}}, \bibinfo{pages}{1246--1254}
  (\bibinfo{year}{2014}).

\bibitem{2001BAR}
\bibinfo{author}{Baroni, S.}, \bibinfo{author}{de~Gironcoli, S.},
  \bibinfo{author}{Dal~Corso, A.} \& \bibinfo{author}{Giannozzi, P.}
\newblock \bibinfo{title}{Phonons and related crystal properties from
  density-functional perturbation theory}.
\newblock \emph{\bibinfo{journal}{Rev. Mod. Phys.}}
  \textbf{\bibinfo{volume}{73}}, \bibinfo{pages}{515} (\bibinfo{year}{2001}).

\bibitem{2011MOM}
\bibinfo{author}{Momma, K.} \& \bibinfo{author}{Izumi, F.}
\newblock \bibinfo{title}{{{\it VESTA3} for three-dimensional visualization of
  crystal, volumetric and morphology data}}.
\newblock \emph{\bibinfo{journal}{Journal of Applied Crystallography}}
  \textbf{\bibinfo{volume}{44}}, \bibinfo{pages}{1272--1276}
  (\bibinfo{year}{2011}).

\bibitem{1999KOK}
\bibinfo{author}{Kokalj, A.}
\newblock \bibinfo{title}{{XCrySDen}---a new program for displaying crystalline
  structures and electron densities}.
\newblock \emph{\bibinfo{journal}{Journal of Molecular Graphics and Modelling}}
  \textbf{\bibinfo{volume}{17}}, \bibinfo{pages}{176--179}
  (\bibinfo{year}{1999}).

\bibitem{2008WIE}
\bibinfo{author}{Wierzbowska, M.}, \bibinfo{author}{de~Gironcoli, S.} \&
  \bibinfo{author}{Giannozzi, P.}
\newblock \bibinfo{title}{Origins of low- and high-pressure discontinuities of
  ${T}_c$ in niobium}.
\newblock \emph{\bibinfo{journal}{eprint arXiv:cond-mat/0504077}}
  (\bibinfo{year}{2006}).

\bibitem{1999MAR}
\bibinfo{author}{Marzari, N.}, \bibinfo{author}{Vanderbilt, D.},
  \bibinfo{author}{De~Vita, A.} \& \bibinfo{author}{Payne, M.~C.}
\newblock \bibinfo{title}{{Thermal Contraction and Disordering of the {Al}(110)
  Surface}}.
\newblock \emph{\bibinfo{journal}{Phys. Rev. Lett.}}
  \textbf{\bibinfo{volume}{82}}, \bibinfo{pages}{3296} (\bibinfo{year}{1999}).
  
\bibitem{2004HON}
\bibinfo{author}{Hongo, K.},
\newblock \bibinfo{title}{Interpretation of Hund's multiplicity rule for the carbon atom.}
\newblock \emph{\bibinfo{journal}{J. Chem. Phys.}}
  \textbf{\bibinfo{volume}{121}}, \bibinfo{pages}{7144--7147} (\bibinfo{year}{2004}).

\bibitem{1959KOH}
\bibinfo{author}{Kohn, W.},
\newblock \bibinfo{title}{Image of the Fermi Surface in the Vibration Spectrum of a Metal}.
\newblock \emph{\bibinfo{journal}{Phys. Rev. Lett.}}
  \textbf{\bibinfo{volume}{2}}, \bibinfo{pages}{393--394} (\bibinfo{year}{1959}).

\bibitem{2015KIM}
\bibinfo{author}{Kim, H.}, \bibinfo{author}{Kang, C.}, \bibinfo{author}{Kim, K.}, 
\bibinfo{author}{Shim, J. H.}, \& \bibinfo{author}{Min, B. I.}
\newblock \bibinfo{title}{Phonon softenings and the charge density wave instability 
in ${R}_{2}{\mathrm{O}}_{2}\mathrm{Sb}$ ($R=\mathrm{rare}\text{-}\mathrm{earth}$ element)}.
\newblock \emph{\bibinfo{journal}{Phys. Rev. B}}
  \textbf{\bibinfo{volume}{91}}, \bibinfo{pages}{165130} (\bibinfo{year}{2015}).

\end{thebibliography}

\begin{thebibliography}{10}
\expandafter\ifx\csname url\endcsname\relax
  \def\url#1{\texttt{#1}}\fi
\expandafter\ifx\csname urlprefix\endcsname\relax\def\urlprefix{URL }\fi
\providecommand{\bibinfo}[2]{#2}
\providecommand{\eprint}[2][]{\url{#2}}

\bibitem{S2013SUE}
\bibinfo{author}{Suetin, D.} \& \bibinfo{author}{Ivanovskii, A.}
\newblock \bibinfo{title}{Electronic properties and fermi surface for new
  {Fe}-free layered pnictide-oxide superconductor {Ba}{Ti}$_2${Bi}$_2${O} from
  first principles}.
\newblock \emph{\bibinfo{journal}{JETP Letters}} \textbf{\bibinfo{volume}{97}},
  \bibinfo{pages}{220--225} (\bibinfo{year}{2013}).

\bibitem{S2013SUE2}
\bibinfo{author}{Suetin, D.} \& \bibinfo{author}{Ivanovskii, A.}
\newblock \bibinfo{title}{Structural, electronic properties, and chemical
  bonding in quaternary layered titanium pnictide-oxides
  {Na}$_2${Ti}$_2$${Pn}_2${O} and {Ba}{Ti}$_2$${Pn}_2${O} ($pn$ = {As}, {Sb})
  from {FLAPW--GGA} calculations}.
\newblock \emph{\bibinfo{journal}{Journal of Alloys and Compounds}}
  \textbf{\bibinfo{volume}{564}}, \bibinfo{pages}{117--124}
  (\bibinfo{year}{2013}).

\bibitem{S2012OUM}
\bibinfo{author}{Ouma, C. N.~M.}, \bibinfo{author}{Mapelu, M.~Z.},
  \bibinfo{author}{Makau, N.~W.}, \bibinfo{author}{Amolo, G.~O.} \&
  \bibinfo{author}{Maezono, R.}
\newblock \bibinfo{title}{Quantum monte carlo study of pressure-induced $b3-b1$
  phase transition in {GaAs}}.
\newblock \emph{\bibinfo{journal}{Phys. Rev. B}} \textbf{\bibinfo{volume}{86}},
  \bibinfo{pages}{104115} (\bibinfo{year}{2012}).
  
\bibitem{S2010WAN}
\bibinfo{author}{Wang, X.~F.} \emph{et~al.}
\newblock \bibinfo{title}{Structure and physical properties for a new layered
  pnictide-oxide: {Ba}{Ti}$_2${As}$_2${O}}.
\newblock \emph{\bibinfo{journal}{Journal of Physics: Condensed Matter}}
  \textbf{\bibinfo{volume}{22}}, \bibinfo{pages}{075702}
  (\bibinfo{year}{2010}).
  
\bibitem{S2012YAJ}
\bibinfo{author}{Yajima, T.} \emph{et~al.}
\newblock \bibinfo{title}{Superconductivity in {Ba}{Ti}$_2${Sb}$_2${O} with a
  $d^1$ square lattice}.
\newblock \emph{\bibinfo{journal}{Journal of the Physical Society of Japan}}
  \textbf{\bibinfo{volume}{81}}, \bibinfo{pages}{103706}
  (\bibinfo{year}{2012}).

\bibitem{S2013YAJ1}
\bibinfo{author}{Yajima, T.} \emph{et~al.}
\newblock \bibinfo{title}{{Synthesis and Physical Properties of the New
  Oxybismuthides {Ba}{Ti}$_2${Bi}$_2${O} and ({SrF})$_2${Ti}$_2${Bi}$_2${O}
  with a $d^1$ Square Net}}.
\newblock \emph{\bibinfo{journal}{Journal of the Physical Society of Japan}}
  \textbf{\bibinfo{volume}{82}}, \bibinfo{pages}{013703}
  (\bibinfo{year}{2013}).

\bibitem{S2009PAO}
\bibinfo{author}{Giannozzi, P.} \emph{et~al.}
\newblock \bibinfo{title}{{QUANTUM ESPRESSO}: a modular and open-source
  software project for quantum simulations of materials}.
\newblock \emph{\bibinfo{journal}{Journal of Physics: Condensed Matter}}
  \textbf{\bibinfo{volume}{21}}, \bibinfo{pages}{395502}
  (\bibinfo{year}{2009}).

\bibitem{S2013SUB}
\bibinfo{author}{Subedi, A.}
\newblock \bibinfo{title}{Electron-phonon superconductivity and charge density
  wave instability in the layered titanium-based pnictide
  {Ba}{Ti}$_{2}${Sb}$_{2}${O}}.
\newblock \emph{\bibinfo{journal}{Phys. Rev. B}} \textbf{\bibinfo{volume}{87}},
  \bibinfo{pages}{054506} (\bibinfo{year}{2013}).

\bibitem{S1957BAR}
\bibinfo{author}{Bardeen, J.}, \bibinfo{author}{Cooper, L.~N.} \&
  \bibinfo{author}{Schrieffer, J.~R.}
\newblock \bibinfo{title}{Theory of superconductivity}.
\newblock \emph{\bibinfo{journal}{Phys. Rev.}} \textbf{\bibinfo{volume}{108}},
  \bibinfo{pages}{1175} (\bibinfo{year}{1957}).

\bibitem{S1965ALL}
\bibinfo{author}{Allen, P.~B.} \& \bibinfo{author}{Dynes, R.~C.}
\newblock \bibinfo{title}{Transition temperature of strong-coupled
  superconductors reanalyzed}.
\newblock \emph{\bibinfo{journal}{Phys. Rev. B}} \textbf{\bibinfo{volume}{12}},
  \bibinfo{pages}{905--922} (\bibinfo{year}{1965}).

\bibitem{S1960ELI}
\bibinfo{author}{Eliashberg, G.~M.}
\newblock \bibinfo{title}{Interactions between electrons and lattice vibrations
  in a superconductor}.
\newblock \emph{\bibinfo{journal}{Sov. Phys.-JETP}}
  \textbf{\bibinfo{volume}{11}}, \bibinfo{pages}{696--702}
  (\bibinfo{year}{1960}).

\bibitem{S2014FRA}
\bibinfo{author}{Frandsen, B.~A.} \emph{et~al.}
\newblock \bibinfo{title}{Intra-unit-cell nematic charge order in the
  titanium-oxypnictide family of superconductors}.
\newblock \emph{\bibinfo{journal}{Nat. Commun.}}
  \textbf{\bibinfo{volume}{5}}, \bibinfo{pages}{5761} (\bibinfo{year}{2014}).
  
\bibitem{S2013YAJ2}
\bibinfo{author}{Yajima, T.} \emph{et~al.}
\newblock \bibinfo{title}{{Two Superconducting Phases in the Isovalent Solid
  Solutions {Ba}{Ti}$_2$${Pn}_2${O} (${Pn}$ = {As}, {Sb}, and {Bi})}}.
\newblock \emph{\bibinfo{journal}{Journal of the Physical Society of Japan}}
  \textbf{\bibinfo{volume}{82}}, \bibinfo{pages}{033705}
  (\bibinfo{year}{2013}).


\end{thebibliography}


\end{document}